\documentclass[twocol]{ametsoc}

\journal{jpo}
\bibpunct{(}{)}{;}{a}{}{,}
\def\pct{\%}
\DeclareMathOperator{\area}{area}%
\DeclareMathOperator{\Id}{Id}%
\DeclareMathOperator{\ii}{i}%
\DeclareMathOperator{\e}{e}%

\newcommand{\re}[1]{\ensuremath{\operatorname{Re}\!{(#1)}}}

\title{Lagrangian geography of the deep Gulf of Mexico}

\authors{P.\ Miron\correspondingauthor{Department of Atmospheric
Sciences, Rosenstiel School of Marine and Atmospheric Science,
University of Miami, Miami, Florida, USA}} \affiliation{Department
of Atmospheric Sciences, Rosenstiel School of Marine and Atmospheric
Science, University of Miami, Miami, Florida, USA}
\email{pmiron@rsmas.miami.edu}

\extraauthor{F.\ J.\ Beron-Vera} \extraaffil{Department of Atmospheric
Sciences, Rosenstiel School of Marine and Atmospheric Science,
University of Miami, Miami, Florida, USA}

\extraauthor{M.\ J.\ Olascoaga} \extraaffil{Department of Ocean
Sciences, Rosenstiel School of Marine and Atmospheric Science,
University of Miami, Miami, Florida, USA}

\extraauthor{G.\ Froyland} \extraaffil{School of Mathematics and
Statistics, University of New South Wales, Sydney, Australia}

\extraauthor{P.\ P\'erez-Brunius} \extraaffil{Departamento de
Oceanograf\'{\i}a F\'{\i}sica, Centro de Investigaci\'on Cient\'{\i}fica
y Educaci\'on Superior de Ensenada, Ensenada, Baja California,
Mexico}

\extraauthor{J.\ Sheinbaum} \extraaffil{Departamento de Oceanograf\'{\i}a
F\'{\i}sica, Centro de Investigaci\'on Cient\'{\i}fica y Educaci\'on
Superior de Ensenada, Ensenada, Baja California, Mexico}

\abstract{Using trajectories from acoustically tracked (RAFOS)
floats in the Gulf of Mexico, we construct a geography of its
Lagrangian circulation within the 1500--2500-m layer.  This is done
by building a Markov-chain representation of the Lagrangian dynamics.
The geography is composed of weakly interacting provinces that
constrain the connectivity at depth.  The main geography includes
two provinces of near equal areas and separated by a roughly
meridional boundary. The residence time is about 4.5 (3.5) years
in the western (eastern) province. The exchange between these
provinces is effected through a slow cyclonic circulation, which
is well constrained in the western basin by preservation of $f/H$,
where $f$ is the Coriolis parameter and $H$ is depth.  Secondary
provinces of varied shapes covering smaller areas are identified
with residence times ranging from about 0.4 to 1.2 years or so.
Except for the main provinces, the deep Lagrangian geography does
not resemble the surface Lagrangian geography recently inferred
from satellite-tracked drifter trajectories. This implies disparate
connectivity characteristics with potential implications for pollutant
(e.g., oil) dispersal at the surface and depth.  A ventilation
conduit through the southeastern corner of the domain from the
Caribbean Sea balanced by weak vertical exchange is also inferred.
This is supported by the inspection of satellite-tracked profiling
(Argo) floats, which, while forming a smaller dataset and having
seemingly different water-following characteristics than the RAFOS
floats, replicate the main aspects of the Lagrangian geography.
Finally, consistency with independent results from a chemical tracer
release experiment is found to provide additional support to the
results.}

\begin{document}

\maketitle

\section{Introduction}

The oil spill produced by the \emph{Deepwater Horizon} drilling rig
explosion in May 2010 \citep{Lubchenco-etal-12} has motivated great
interest in the Lagrangian circulation of the Gulf of Mexico (GoM).
This is reflected in the execution in recent years of a number of
field campaigns dedicated to observe its surface Lagrangian
circulation.  A main reason for investigating the surface Lagrangian
circulation is found in the very tangible effects it had on the
evolution of the oil slick that emerged from the ocean floor
\citep{Olascoaga-Haller-12}.  The main campaigns have been the Grand
LAgrangian Deployment (GLAD) in July 2012 \citep{Olascoaga-etal-13,
Poje-etal-14, Beron-LaCasce-16} and the LAgrangian Submesoscale
ExpeRiment (LASER) in February 2016 \citep{Miron-etal-17,
Novelli-etal-17}.  These two campaigns contributed to nearly duplicate
the satellite-tracked surface drifter database existing prior to
the oil spill, which consisted mainly of drifter trajectories from
the National Oceanic and Atmospheric Administration (NOAA) Global
Drifter Program \citep[GDP,][]{Lumpkin-Pazos-07} and the Surface
Current Lagrangian-Drifter Program \citep[SCULP,][]{Sturgers-etal-01,
Ohlmann-Niiler-05}, see \citet{Miron-etal-17} for details.

Large amounts of oil were reported to stay submerged and to persist
for months without substantial biodegradation \citep{Camilli-etal-10}.
Yet the effects that the deep Lagrangian circulation had on the
submerged oil remained elusive, which directly or indirectly motivated
the execution of experiments to also observe the Lagrangian circulation
at depth.  One experiment consisted in a deployment of acoustically
tracked floats in a mission that started in 2011 and lasted out to
2015 \citep{Hamilton-etal-16}.  This contributed to augment the
existing submerged float database, consisting mainly of profiling
floats from a dedicated experiment \citep{Weatherly-etal-05} and
routine sensing of the deep global ocean \citep{Roemmich-etal-09}.
Another experiment involved the release of a chemical tracer at
depth in July 2012 near the \emph{Deepwater Horizon} site and its
subsequent sampling over the course of one year \citep{Ledwell-etal-16}.

An aspect of the deep Lagrangian circulation highlighted by the
dedicated profiling float experiment \citep{Weatherly-etal-05} was
the restricted communication between the eastern and western GoM
basins and also a cyclonic circulation at about 900 m in the
southwestern sector.  Analysis of the acoustically tracked float
trajectories in the western basin \citep{Perez-etal-17} from the
recent experiment \citep{Hamilton-etal-16} further revealed the
existence of a cyclonic boundary current below 900 m and a cyclonic
gyre in the abyssal plain consistent with numerical studies
\citep{Oey-Lee-02}, and the analysis of hydrographic data
\citep{DeHaan-Sturges-05} and deep-water moorings \citep{Tenreiro-etal-17}.
Direct inspection of the same acoustically tracked float trajectories
\citep{Perez-etal-17}, as well as rough estimates of connectivity
between the eastern and western basins \citep{Hamilton-etal-16},
suggests that the exchange between them occurs along the boundary
following a cyclonic circulatory motion. In turn, the analysis of
the dispersion of the chemical tracer released at depth in the
eastern basin \citep{Ledwell-etal-16} concluded that homogenization
by stirring and mixing is substantially faster in the GoM than in
the open ocean.  The main source of energy in the deep eastern basin
is presumably provided by the Loop Current by inducing a deep flow
through baroclinic instabilities, deep eddies, and topographic
Rossby waves which can transfer energy toward the western basin
\citep{Sheinbaum-etal-16, Hamilton-etal-16, Donohue-etal-16}.

The goal of this paper is to shed new light on the deep Lagrangian
circulation in the GoM by using probabilistic tools from nonlinear
dynamical systems.  These are applied on the above acoustically
tracked float trajectories with a focus on connectivity.  Investigating
connectivity with the probabilistic nonlinear dynamics tools boils
down to analyzing the eigenvectors of a transfer operator approximated
by a matrix of probabilities of transitioning between boxes of a
grid, which provides a discrete representation of the Lagrangian
dynamics \citep{Froyland-etal-14}.  Markov-chain representations
of this type had originally been used to approximate almost-invariant
sets in nonlinear dynamical systems using short-run trajectories
\citep{Hsu-87, Dellnitz-Junge-99, Froyland-05}, and in the ocean
context to determine the extent of Antarctic gyres in 2-
\citep{Froyland-etal-07} and 3-space \citep{Dellnitz-etal-09}
dimensions.  This eigenvector method has been recently applied on
drifter data to construct a geography of the surface Lagrangian
circulation \citep{Miron-etal-17}.  A Lagrangian geography is
composed of dynamical provinces that delineate weakly interacting
basins of attraction for almost-invariant attractors, which imposes
constraints on connectivity.  Here we construct a geography for the
deep Lagrangian circulation, providing firm support to earlier
inferences from the direct inspection float trajectories and,
furthermore, revealing a number of aspects transparent to traditional
Lagrangian data analysis.

\section{Data}

\begin{figure*}[t]
  \centering%
  \includegraphics[width=\textwidth]{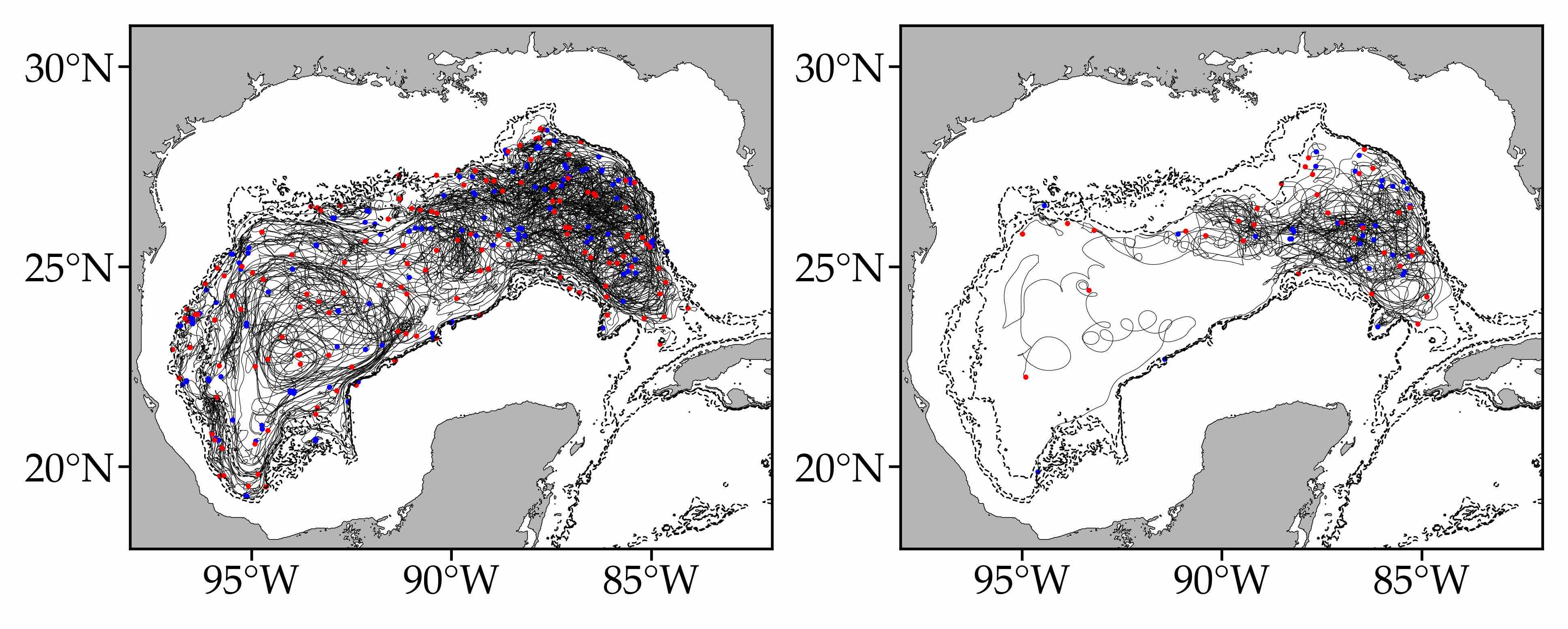}%
  \caption{Trajectories of RAFOS floats ballasted at 1500 (left)
  and 2500 (right) m in the Gulf of Mexico over 2011--2015. Indicated
  are initial (blue dots) and final (red dots) positions of the
  floats.  The dash lines indicate, from outside to inside, the
  1500-, 1750-, and 2500-m isobaths.}
  \label{fig:rafos}%
\end{figure*}

The main dataset analyzed in this paper is composed of trajectories
produced by a total of 154 quasi-isobaric acoustically tracked RAFOS
(SOund Fixing And Ranging or SOFAR, spelled backward) floats
\citep{Rossby-etal-86} deployed in the GoM \citep{Hamilton-etal-16}.
Starting in 2011, 121 floats ballasted for 1500 m and 31 floats for
a lower depth of 2500 m were deployed in the following 2 years.
Each float recorded position fixes 3 times daily, with record lengths
varying between 7 days and 1.5 years.  This sampling is too frequent
for an estimated uncertainty of the order of 5 km, so we here
consider daily interpolated trajectories. The float deployment
during the first 2 years of a 4-year-long program was performed by
several U.S.\ (Woods Hole Oceanographic Institution, Leidos
Corporation, University of Colorado) and Mexican (Centro de
Investigaci\'on Cient\'{\i}fica y de Educaci\'on Superior de Ensenada)
teams sponsored by the U.S.\ Bureau of Ocean Energy Management
(BOEM).  The records of the last floats deployed ended in summer
2015.  The recorded trajectories of all floats are shown in Fig.\
\ref{fig:rafos} (deployment locations and final positions are
indicated in blue and red, respectively).  The trajectories cover
a region bounded for the most part by the 1750-m isobath (dashed
lines in Fig.\ \ref{fig:rafos} indicate, from outside to inside,
the 1500-, 1750-, and 2500-m isobaths).  Note that while the floats
are too deep to escape the GoM through the Straits of Florida (where
the maximum depth is roughly 700 m), they are capable of escaping
through the Yucatan Channel (where the maximum depth is about 2000
m).  Mooring measurements suggest that the latter is indeed possible
as they have revealed a countercurrent between 500 and 1750 m on
the western and eastern sides of the Yucatan Channel
\citep{Sheinbaum-etal-02}.  However, no float is seen to travel
into the Caribbean Sea.  Confinement of the 1500- and 1750-m floats
within the region bounded by the 1750-m isobath suggests predominantly
columnar motion.  This is confirmed by the analysis presented below,
which ignores the depth of the floats to maximize the number of
available trajectories.

We also analyze trajectories recorded by all available (60) profiling
floats in the GoM from the Argo Program \citep{Roemmich-etal-09}.
Unlike RAFOS float positions, Argo float positions are recorded
every 10 days after the float descends down to a parking depth of
1000 m, where it drifts for 9 days, and further to 2000 m to begin
profiling temperature and salinity in its ascend back to the surface.
The trajectories of the Argo floats roughly sample the same area
as the RAFOS floats albeit much less densely.  However, the way
that the Argo floats sample the deep Lagrangian circulation may be
expected to differ from the way done it by the RAFOS floats, which
remain at all times parked at a fixed depth.  Despite this expectation,
we show that Argo floats replicate some important aspects of the
deep circulation inferred using the RAFOS floats.

A third set of independent data considered is composed of concentrations
of chemical tracer from a release experiment \citep{Ledwell-etal-16}.
In the experiment, a 25-km-long streak of CF$_3$SF$_5$ was injected
on an isopycnal surface about 1100-m deep and 150 m above the bottom,
along the continental slope of the northern GoM, about 100-km
southwest of the \emph{Deepwater Horizon} oil well, where oil was
detected at depth after its explosion.  The tracer was sampled
between 5 and 12 days after release, and again 4 and 12 months after
release.

\section{Theory}

\subsection{Transfer operator and transition matrix}

Let $X \in \mathbb{R}^2$ be a closed flow domain on the plane.  We
assume that the Lagrangian dynamics is governed by an advection--diffusion
process.  A tracer initialized at position $y\in X$ (represented
by a delta-measure $\delta_y$) therefore evolves passively $T > 0$
units of time to a probability density $K(x,y)$, with $x\in X$.
The function $K(x,y)$ is clearly nonnegative, and we normalize in
$x$ so that
\begin{equation}
\label{densityK}
\smash{\int_X K(x,y)\,dx = 1}.
\end{equation}
The function $K(x,y)$
is a (bounded) stochastic kernel; cf., e.g., Sections 5.7 and 11.7
of \citet{Lasota-Mackey-94} for a discussion of
stochastic kernels and advection--diffusion equations, respectively.
To evolve a general initial density $h : X\to \mathbb{R}^+$ forward
$T$ units of time, we define a Markov operator, known as the
Perron-Frobenius operator, or more generally a \emph{transfer
operator}, $\mathcal{P}:L^1(X)\circlearrowleft$, as
\begin{equation}
  \mathcal P h(x) = \int_X K(x,y)h(y)\,dy.
  \label{eq:PF}
\end{equation}
The density $\mathcal Ph(x)$ is the result of evolving $h(x)$
forward $T$ units of time under the advection--diffusion dynamics.

Note that the only time dependence is the duration of time $T$.  In
particular, we do not model variation of the advection--diffusion
dynamics as a function of initial time.  This is appropriate for a
probabilistic description of the dynamics, as done in statistically
stationary turbulence \citep{Orszag-77}, yet it is also a consequence
of the nature of the dataset considered here.  In either case the
significance of the time homogeneity assumption can only be assessed
a posteriori, as we do here.

A discretization of the transfer operator can be attained using a
Galerkin approximation referred to as Ulam's method \citep{Ulam-79,
Kovacs-Tel-89}.  This involves partitioning the domain $X$ into a
grid of $N$ connected boxes $\smash{\{B_1, \dotsc, B_N\}}$ and
projecting functions in $L^1(X)$ onto the finite-dimensional
approximation space $\smash{V_N := \mbox{span}\{\mathbf 1_{B_1}(x),
\dotsc ,\mathbf 1_{B_N}(x)\}}$, where $\mathbf 1_{B_i}(x) = 1$ for
$x \in B_i$ and 0 otherwise is the indicator function of set $B_i$.
Define $\pi_N:L^1(X) \to V_N$ by
\begin{equation}
  \smash{\pi_Nh(x)} = \sum_{j=1}^N \frac{\int_{B_j}
  h(x)\,dx}{\area(B_j)}\, \mathbf 1_{B_j}(x).
\end{equation}
To calculate the projected action of $\mathcal{P}$ on $V_N$ we compute:
\begin{eqnarray}
   \pi_N\mathcal P\mathbf 1_{B_i}(x)
	\!\!\!\!&=&\!\!\!\!%
	\sum_{j=1}^N
	\frac{\int_{B_j}\mathcal P\mathbf 1_{B_i}(x)\,dx}{\area(B_j)}\,
	\mathbf 1_{B_j}(x)\nonumber\\
   &=&\!\!\!\!%
	\sum_{j=1}^N\frac{\int_{B_j} \int_X
	K(x,y)\mathbf 1_{B_i}(y)\,dydx}{\area(B_j)}\,
	\mathbf 1_{B_j}(x)\nonumber\\
	&=&\!\!\!\!%
	\sum_{j=1}^N\underbrace{\frac{\int_{B_j}\int_{B_i}
	K(x,y)\,dydx}{\area(B_j)} }_{=:P_{ij}}\, \mathbf 1_{B_j}(x)
\end{eqnarray}
The $(i,j)$-th entry of the array $P_{ij}$ represents an approximation
of the kernel $K(x,y)$ for $y\in B_i, x\in B_j$.  We assume from
now on that the grid is regular, i.e., $\area(B_i) = \area(B_j)$
for all $1\le i,j\le N$.  Because $K(x,y)$ represents the density
obtained by evolving $\delta_y$ forward $T$ units of time,
\begin{equation}
  \label{eq:P} P_{ij} = \frac{\int_{B_j}\int_{B_i}
  K(x,y)\,dydx}{\area(B_i)}
\end{equation}
is the proportion of tracer beginning in $B_i$ that is found in
$B_j$ after $T$ units of time (``proportion'' because of integration
over $B_j$ and the property (\ref{densityK})), averaged over the
initial tracer in $B_i$ (``averaged'' because of the integration
over $B_i$ and division by $\area(B_i)$).

If we are presented with tracer data in the form of trajectories
of individual tracer particles, by considering a sufficiently large
number of particles over a total time horizon $[0, \mathcal T]$ we
can estimate the entries of $P_{ij}$ as
\begin{equation}
  P_{ij} \approx \frac{\mbox{\# $x\in B_i$
  at $t\in [0,\mathcal T - T]$ and $B_j$ at
  $t+T$}}{\mbox{\# $x\in B_i$ at
  $t\in [0,\mathcal T - T]$}}.
  \label{eq:Papprox}
\end{equation}
The \emph{transition matrix} $\smash{P \in \mathbb{R}^{N \times
N}}$ defines a Markov-chain representation of the dynamics, with
the entries $P_{ij}$ equal to the conditional transition probabilities
between boxes, which are represented by the states of the chain.

The forward evolution of the discrete representation of $h(x)$,
$\mathbf h = (h_1\, \cdots \, h_N)$, is calculated under left
multiplication, i.e.,
\begin{equation}
  \mathbf h^{(k)} = \mathbf h P^{k},\quad k = 1, 2, \dotsc.
  \label{eq:pP}
\end{equation}

We note that the discrete evolution described by $P$ introduces
additional diffusion with magnitude of the order of the box diameters
\citep{Froyland-13}.

\subsection{Ergodicity, mixing, attracting sets, residence time,
and retention time}

Because the transition matrix $P$ is row stochastic, i.e.,
$\smash{\sum_{j=1}^N P_{ij}} = 1 \ \forall i$, $\mathbf{1} = (1
\dotsb 1)$ is a \emph{right} eigenvector with eigenvalue $\lambda
= 1$, i.e., $P\mathbf{1} = \mathbf{1}$.  The eigenvalue $\lambda =
1$ equals the spectral radius of $P$.  The associated nonunique
\emph{left} eigenvector $p$ is normalized to a probability vector
($\smash{\sum_{i=1}^N p_i = 1}$) which is invariant (because $pP =
p$) and nonnegative \citep[by the Perron--Frobenius
theorem,][]{Horn-Johnson-90}.

We call $P$ \emph{irreducible} (or ergodic) if for each $1 \le i,j
\le N$ there is an $n_{ij} < \infty$ such that $\smash{(P^{n_{ij}})_{ij}
> 0}$.  All states of an irreducible Markov chain communicate, the
eigenvalue $\lambda = 1$ is simple, and the corresponding left
eigenvector $p$ is strictly positive \citep{Horn-Johnson-90}.  We
call $P$ \emph{aperiodic} (or mixing) if there is an $i$ such that
$\gcd\{n\ge 0 \mid \smash{(P^{n})_{ii} > 0}\} = 1$.  For aperiodic
$P$ one has $\smash{p = \lim_{k\to\infty} \mathbf h P^k}$ for
\emph{any} initial probability vector $\mathbf h$.

Suppose that $P$ is irreducible on some class of states $S \subset
\{1,\dotsc,N\}$.  We call $S$ an \emph{absorbing closed communicating
class} if $P_{ij} = 0$ for all $i \in S, j\notin S$, and $P_{ij} >
0$ for some $i \notin S, j\in S$;  cf.\ \citet{Froyland-etal-14}.
The set $\smash{B_S = \bigcup_{i\in S} B_i\subset X}$ forms an
approximate time-asymptotic forward-invariant attracting set for
trajectories starting in $\smash{X = \bigcup_{i=1}^N B_i}$.

Markov chain theory also provides a very simple means of determining
the mean residence time in a collection of boxes.  For $A\subset
X$ let $P|_A$ be the transition matrix $P$ restricted to the subset
of indices corresponding to boxes whose union is $A$.  The mean
time $\tau_i$ of a trajectory initialized in box $B_i$ to move out
of $A$, also known as the mean time to hit the complement of $A$,
is given by the solution of the linear equation (cf., e.g.,
\citet{Norris-98} and \citet{Dellnitz-etal-09} for its use in the
context of ocean dynamics):
\begin{equation}
  (\Id - P|_A)\tau/T = \mathbf{1},
  \label{eq:res}
\end{equation}
where $\Id$ is the identity matrix. The mean value of $\tau$ within
$A$ is a measure of the residence time of the entire set.

Another timescale related to residence time is the time-asymptotic
retention time.  As before, let $A\subsetneq X$ be the set for which
we wish to allocate a retention time and denote $P|_A$ the restriction
of $P$ to boxes whose union is $A$.  If $P$ is mixing, the leading
positive eigenvalue of $P|_A$, $\lambda_A$, is strictly less than
1.  If one conditions on the fact that trajectories have already
remained in $A$ sufficiently long (so they are distributed like
$p_A\ge 0$, the leading left eigenvector of $P|_A$), then the
probability of remaining in $A$ for one more application of $P$ is
$\lambda_A$.  The probability of a point in $A$ (distributed according
to $p_A$) remaining in $A$ for $k\ge 1$ applications of $P$ is
\begin{equation}
  \frac{\lambda_A^k}{\sum_{k=1}^\infty \lambda_A^k} =
  \frac{\lambda_A^k}{\lambda_A/(1-\lambda_A)} =
  (1-\lambda_A)\lambda_A^{k-1}.
\end{equation}
Hence, the mean retention time $\mathfrak T$ for points in $A$
distributed according to $p_A$ is
\begin{eqnarray}
    \mathfrak T/T &=& \sum_{k=1}^\infty
    \left((1-\lambda_A)\lambda_A^{k-1}\right)\cdot k\nonumber\\ &=&
    (1-\lambda_A)\sum_{k=1}^\infty k\lambda_A^{k-1}\nonumber\\ &=&
    (1-\lambda_A)\cdot \frac{1}{(1-\lambda_A)^2}\nonumber\\ &=&
     \frac{1}{1-\lambda_A}.  \label{eq:ret}
 \end{eqnarray}
Note that if the mean value of $\tau$, $\tau_A$, is computed according
to $p_A$, i.e., $\tau_A = \sum_{j\in A} (p_A)_j\tau_j$, then $\tau_A
= \mathfrak T$.

\subsection{Lagrangian geography from almost-invariant decomposition}

Revealing those regions in which trajectories tend to stay for a
long time before entering another region is key to assessing
connectivity in a flow.  Such forward time-asymptotic
\emph{almost}-invariant sets and their corresponding backward-time
basins of attraction can be framed \citep{Froyland-etal-14} by
inspecting eigenvectors of $P$ with $\lambda \approx 1$.

The magnitude of the eigenvalues quantify the geometric rates at
which eigenvectors decay.  Those left eigenvectors with $\lambda$
closest to 1 are the slowest to decay and thus represent the most
long-lived transient modes \citep{Froyland-97, Pikovsky-Popovych-03}.
For a given $\lambda \approx 1$, a forward time-asymptotic
almost-invariant set will be identified with the support of similarly
valued and like-sign elements in the left eigenvector.  Regions
where the magnitude of the left eigenvector is greatest are the
most dynamically disconnected and take the longest times to transit
to other almost-invariant sets.

The multiple backward-time basins of attraction are identified by
boxes where the corresponding right eigenvectors take approximately
constant values \citep[cf.][for the simpler single basin case]{Koltai-11}.
Decomposition of the ocean flow into weakly disjoint basins of
attraction for time-asymptotic almost-invariant attracting sets
using the above eigenvector method has been shown \citep{Froyland-etal-14,
Miron-etal-17} to form the basis of a \emph{Lagrangian geography}
of the ocean, where the boundaries between basins are determined
from the Lagrangian circulation itself, rather than from arbitrary
geographical divisions.

We note that the eigenvector method differs from the flow network
approach \citep{Rosi-etal-14, Sergiacomi-etal-15}. The eigenvector
method analyzes time-asymptotic aspects of the dynamics through
spectral information from the generating Markov chain, while the
flow network approach computes various graph-based quantities for
finite-time durations to study flow dynamics.

\section{Results}

\subsection{Building a Markov-chain model}

\begin{figure*}[t]
  \centering%
  \includegraphics[width=\textwidth]{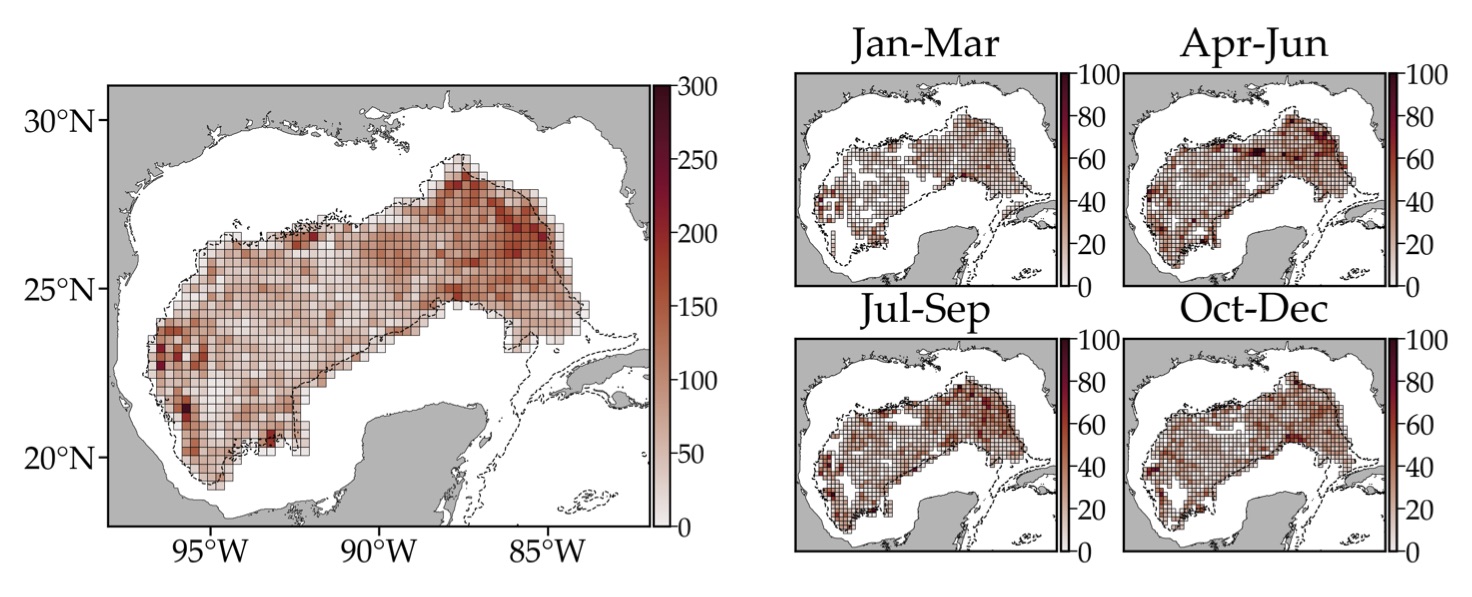}%
  \caption{For daily interpolated trajectories, number of RAFOS
  floats per box in the grid into which the domain visited has been
  discretized, independent of time over the entire 2011--2015 period
  (left) and within each season in this period (right). The dashed
  line, here and all the subsequent figures, represent the  1750-m
  isobath.}
  \label{fig:density}%
\end{figure*}

To discretize the deep-ocean Lagrangian dynamics in the GoM, we
laid down on the region $X$ spanned by the trajectories of the RAFOS
floats in Fig.\ \ref{fig:rafos} a grid with $N = 946$ boxes of
roughly 25-km a side. The size of the boxes was selected to maximize
the grid's resolution while each individual box is sampled by enough
trajectories (recall that the float's position is determined with
a precision of about 5 km, so the uncertainty area around a float's
location is roughly 8 times smaller than the area of a box in the
grid).  Figure \ref{fig:density} shows number of floats per box in
the grid, independent of time over the entire 2011--2015 period
(left) and within each season in this period (right). Regions not
visited by floats are found in each season, particularly in winter.
Ignoring time, there are on average 86 floats per box, with 1 box
having as many as 286 floats and 7 boxes only 1 float. Overall,
while the float coverage may not be dense enough to carry out a
seasonal analysis, it is sufficient in space to build a Markov-chain
model that assumes time homogeneity \citep{Maximenko-etal-12,
vanSebille-etal-12, Miron-etal-17, McAdam-vanSebille-18}.

Before getting into the specifics of the computation of the transition
matrix $P$ defining the Markov chain, it is important to note that
formula \eqref{eq:P} for $P$ does not require the number of
trajectories that sample the boxes in the partition to be equal for
each box.  Thus the nonuniform sampling of the RAFOS floats is not
an impediment for computing $P$.  Nevertheless, to make sure that
this does not introduce any biases in our analysis, we have also
computed various $P$s by randomly choosing a fixed number (50) of
trajectories.  This required us to eliminate 94 boxes from the
original partition.  The resulting $P$s were found to produce results
that could not be distinguished from those produced by the $P$
computed using all available trajectories, as done as follows.

Using formula \eqref{eq:Papprox}, we computed the $P_{ij}$ entry
of the transition matrix by counting the number of floats that
starting in box $B_i$ on any day landed in box $B_j$ after $T = 7$
days within the entire record of float data, which lasts $\mathcal
T \approx 4$ years.  A transition time $T = 7$ days in general
guarantees interbox communication.  Furthermore, $T = 7$ days is
larger than the Lagrangian decorrelation timescale, which we estimated
to be of about 5 days for half decorrelation consistent with earlier
estimates \citep[cf.][]{LaCasce-08}.  Markovian dynamics can
be expected to approximately hold as there is negligible memory
farther than 7 days into the past.  A similar reasoning was applied
in applications involving surface drifters \citep{Maximenko-etal-12,
vanSebille-etal-12, Miron-etal-17, McAdam-vanSebille-18}, in which
case the transition time was taken shorter due to the shorter
decorrelation time near the ocean surface \citep{LaCasce-08}.  The
results presented below were found insensitive to variations of $T$
in the range 7--21 days.

\subsection{Assessing communication within the Markov chain}

A Markov chain can be seen as a directed graph with vertices
corresponding to states in the chain, and directed arcs corresponding
to one-step transitions of positive probability.  This allows one
to apply Tarjan's algorithm \citep{Tarjan-72} to assess communication
within a chain.  Specifically, the Tarjan algorithm takes such a
graph as input and produces a partition of the graph's vertices
into the graph's strongly connected components.  A directed graph
is strongly connected if there is a path between all pairs of
vertices.  A strongly connected component of a directed graph is a
maximal strongly connected subgraph and by definition also a maximal
communicating class of the underlying Markov chain.

\begin{figure}[t]
  \centering%
  \includegraphics[width=\linewidth]{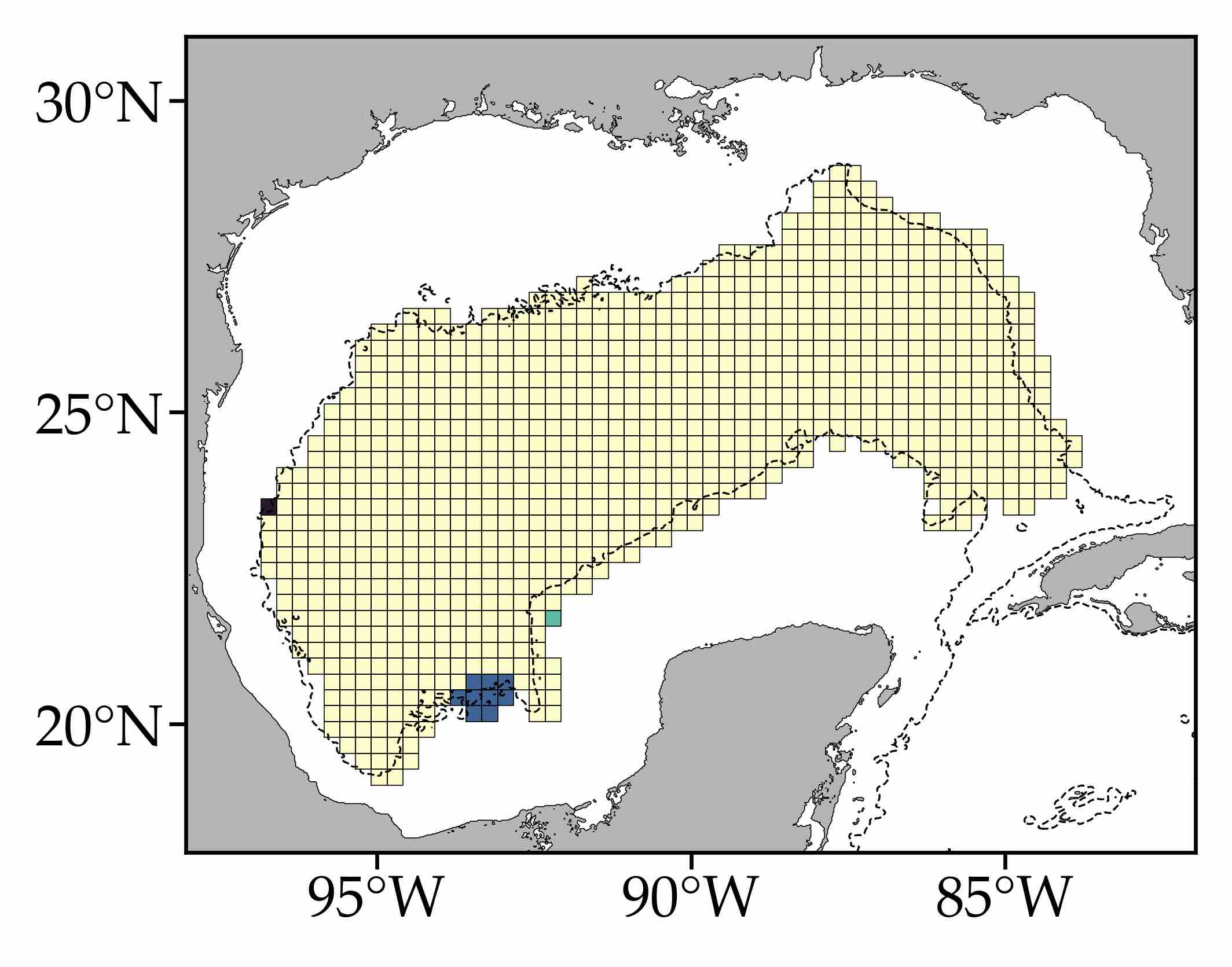}%
  \caption{Grouping of the states of the Markov chain associated
  with the matrix $P$ of transition probabilities of RAFOS floats
  moving over 7 days between boxes of the partition of the domain
  spanned by their trajectories into classes according to their
  communication type.  Boxes indicated with the same color belong
  to the same class. There are 4 maximal communicating classes, with
  a large class (yellow) encompassing most of the states in the
  Markov chain (boxes of the partition) which is absorbing.}
  \label{fig:classes}%
\end{figure}

Applying the Tarjan algorithm to the directed graph associated with
the Markov chain derived using the float trajectory data, we found
a total of four maximal communicating classes. Each one of these
classes is indicated with a different color in Fig.\ \ref{fig:classes}.
One class, denoted $S_\mathrm{L}$, is large, composed of the majority
of the states in the chain or boxes of the partition (935 out of a
total of $N = 946$). From direct computation, $P_{ij} = 0$ for all
$i \in S_\mathrm{L}$, $j \notin S_\mathrm{L}$, i.e., $S_\mathrm{L}$
is closed. The classes in the complement of $S_\mathrm{L}$ are
small, with 2 formed by a single state and another one formed by 9
states.  Direct computation shows that $(P^m)_{ij} > 0$ for some
$m$ for all $i \notin S_\mathrm{L}$, $j \notin S_\mathrm{L}$.  This
reveals that $S_\mathrm{L}$ is absorbing.  Because $S_\mathrm{L}$
is communicating and closed, $P$ restricted to $S_\mathrm{L}$, i.e.,
$P|_{S_\mathrm{L}}$, is irreducible and provided that no state
repeatedly occurs, it will be mixing, i.e., a unique, limiting
invariant probability density will be supported on $S_\mathrm{L}$.
Because $S_\mathrm{L}$ is absorbing, its small complement can be
safely ignored from the analysis without affecting the results by
replacing $P$ with $P|_{S_\mathrm{L}}$.

\subsection{Forward evolution of the probability density of a
tracer}

\begin{figure*}[t]
  \centering%
  \includegraphics[width=\textwidth]{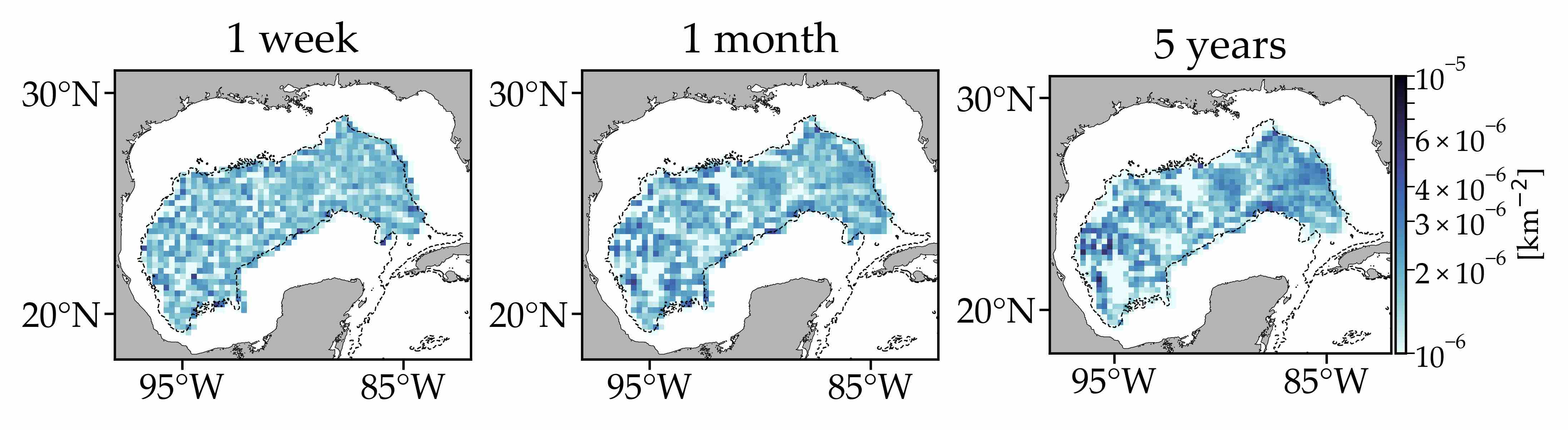}%
  \caption{Selected snapshots of the evolution of an initially
  uniformly distributed tracer probability density under the action
  of the underlying flow map, whose discrete representation is
  provided by the float-data-derived transition matrix $P$.}
  \label{fig:limit}%
\end{figure*}

Figure \ref{fig:limit} shows selected snapshots of the push forward
of the probability density of a tracer initially uniformly distributed
within the layer between 1500 and 2500 m in the GoM under the
discrete action of the underlying flow map.  At the coarse-grained
level given by the grid defined above, this is defined by \eqref{eq:pP}
with $\mathbf h = \mathbf 1/N$ and $P$ as derived using the float
data.  Because we used 7-day-long trajectory pieces to construct
$P$, one application of $P$ is equivalent to evolving in forward
time for $T = 7$ d.  Note that the density eventually settles on a
nonuniform distribution which appears to be invariant.

The regions where the density distribution of Fig.\ \ref{fig:limit}
locally maximizes represent vertical ``outwelling'' sites.  Likewise,
there is vertical ``inwelling'' in the regions where the density
locally minimizes.   Volume conservation in either case implies
vertical motion.  While the direction of this motion cannot be
determined from the analysis of the float trajectories on a single
layer, some sense may be made of its magnitude by comparing area
change estimates obtained using the deep floats and the satellite-tracked
surface drifter data employed in \citet{Miron-etal-17}.

\begin{figure}[t]
  \centering%
  \includegraphics[width=\linewidth]{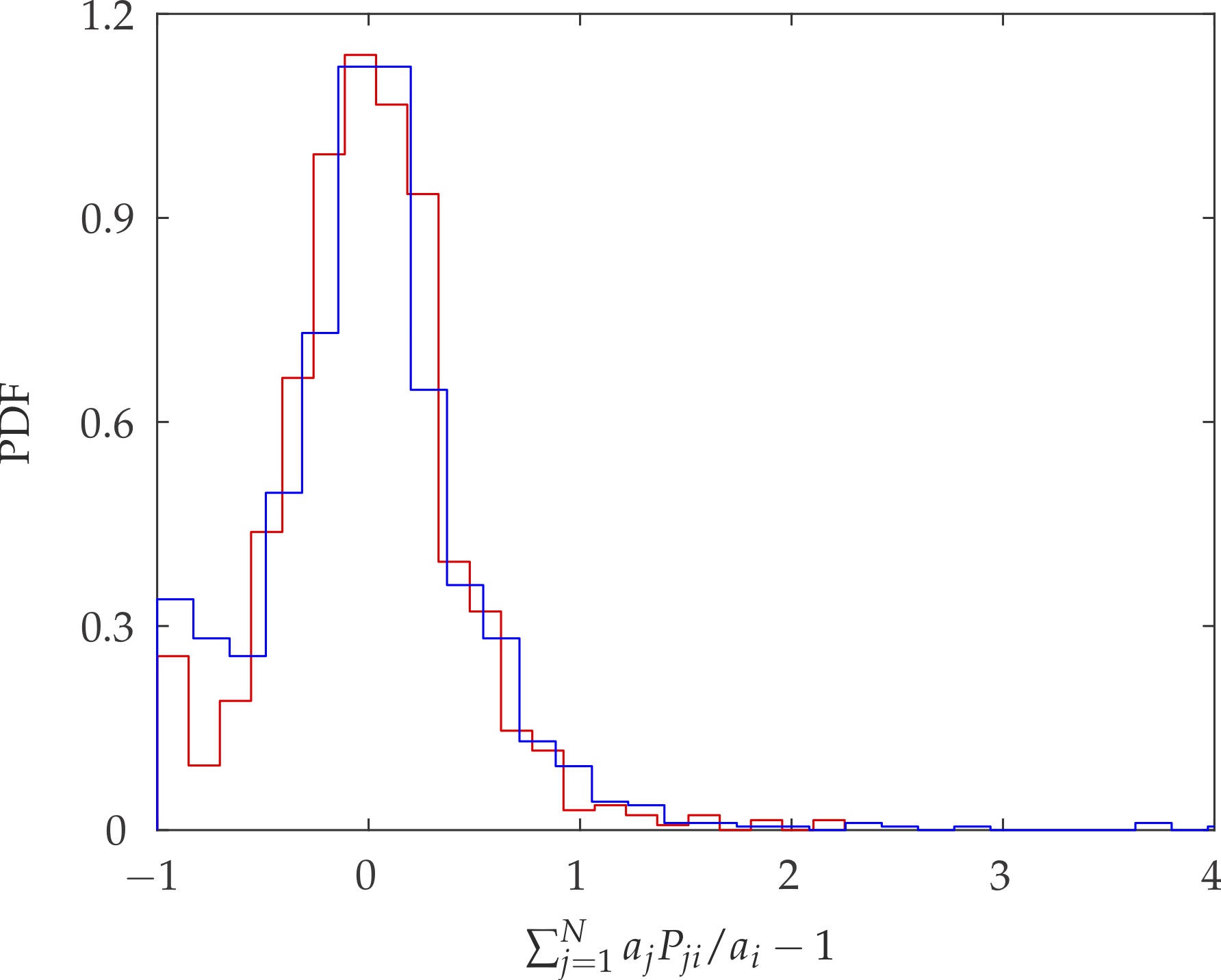}%
  \caption{Probability density function estimates of relative area
  change computed over 7 days using the RAFOS deep float data (red)
  and satellite-track surface drifter data (blue).}
  \label{fig:vertical}%
\end{figure}

Let $a_i$ be the area of box $B_i$ of the domain on which the float
trajectories lie.  After 1 application of $P$, equivalently 7 d,
we have $a'_i := \smash{\sum_{j=1}^N a_j P_{ji}}$.  If the flow
were area-preserving, we would expect $a'_i = a_i$.  At depth, area
dispersion ($a'_i > a_i$) corresponds to vertical inwelling, while
area concentration ($a'_i < a_i$) to vertical outwelling.  At the
surface, area dispersion and concentration correspond to up- and
downwelling, respectively.  In Fig.\ \ref{fig:vertical} we show
probability density function (PDF) estimates of relative area change
$a'_i/a_i - 1 \ge -1$ inferred using deep float data (red) and
surface drifter data (blue).  The surface relative area change is
computed over 7 days and using a partition into boxes of similar
size as at depth, approximately 625 km$^2$.  Note that both PDFs
peak at 0, with the drifter PDF showing longer tails toward large
positive values and higher negative values than the float PDF.
Overall area preservation is seen to dominate equally at the surface
and depth, while a larger tendency to disperse and concentrate areas
is observed at the surface than at depth.

\subsection{Analysis of the Markov chain's eigenspectrum}

We now proceed to determine the level of connectivity within the
horizontal domain in the layer visited by the deep floats by applying
the eigenvector method on the matrix $P$.

\begin{figure}[t]
  \centering%
  \includegraphics[width=\linewidth]{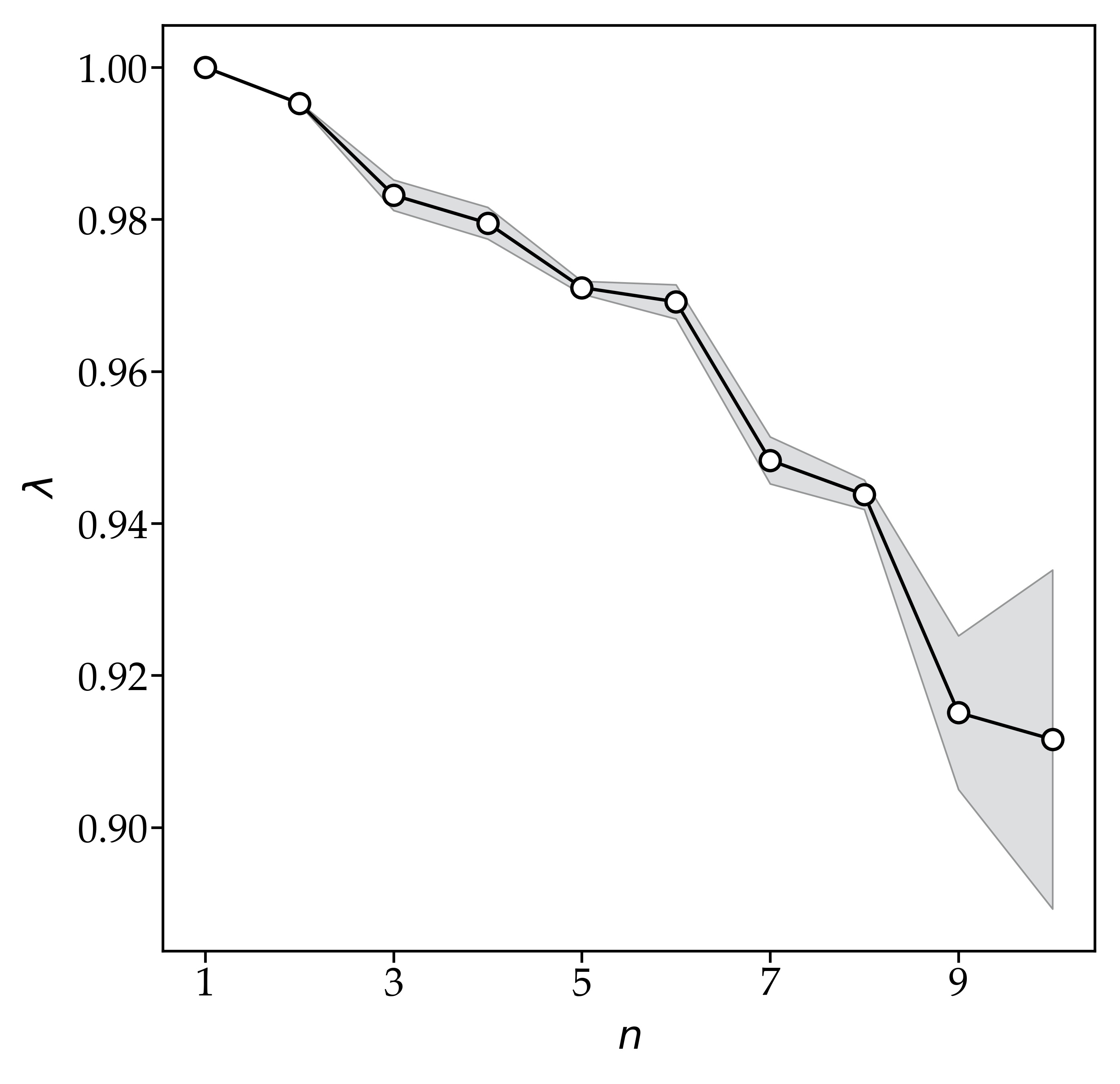}
  \caption{A portion of the discrete eigenspectrum of the transition
   matrix $P$ showing the top 10 real eigenvalues (circles) and
   uncertainties (gray shade) representing the median absolute
   deviation about them in ensemble of eigenvalues computed using
   transition matrices produced by randomly perturbing the float
   trajectories.}
  \label{fig:eigenvalues}%
\end{figure}

The eigenspectrum of $P$ is composed of a total of $N = 946$
eigenvalues.  As noted above, only eigenvectors of $P$ corresponding
to eigenvalues close to unity are relevant, and because $P$ is very
sparse, one may use Lanczos-type methods \cite{Lehoucq-etal-98} to
compute the $n\ll N$ largest eigenvalues.  With this in mind, we
show in Fig.\ \ref{fig:eigenvalues} the eigenspectrum of $P$ restricted
to $\lambda > 0.9$.  This top 10\pct\, portion of the eigenspectrum
includes 10 eigenvalues, yet not all 10 associated eigenvectors
might need to be taken into account.  Indeed, thinking of $\lambda_n
< 1$ as a decay rate for a signed density revealed by the $n$th
left eigenvector of $P$, if a large spectral gap between two
collections of eigenvalues is present, then the densities revealed
by the eigenvectors associated with eigenvalues prior to the gap
will decay much more slowly and survive over much longer timescales
than the densities revealed by the eigenvectors associated with the
eigenvalues after the gap.  Therefore, the presence of a pronounced
eigengap provides rationale for stopping eigenvector analysis.
However, inspection of Fig.\ \ref{fig:eigenvalues} does not reveal
any gap that strongly suggests a cutoff for analysis except, perhaps,
at $n = 8$ or $n = 6$.  Yet only the gap at $n = 6$ may be considered
significant with respect to the uncertainty of the eigenvalue
computation.  The gray shade in Fig.\ \ref{fig:eigenvalues} represents
an uncertainty of the computation of $\lambda_n$ as measured by
the median absolute deviation about $\lambda_n$ in an ensemble of
1000 realizations computed from transition matrices constructed
using randomly perturbed float trajectories with 5 km amplitude,
corresponding to the float positioning accuracy.  This uncertainty
measure grows noticeably beyond $n = 8$, making the gap there not
significant.  This suggests, in the absence of evidence establishing
any better criterion, that the analysis should not be extended
beyond the 6th eigenvector.  It must be noted that we have so far
implicitly referred to the real eigenspectrum of $P$.  Interspersed
among the dominant eigenvalues (in absolute magnitude) are
complex-conjugate eigenvalue pairs.  We will discuss a relevant
complex eigenvector pair after discussing the real eigenvectors,
restricted to the first 6 as just articulated.

\begin{figure*}[t]
  \centering%
  \includegraphics[width=\textwidth]{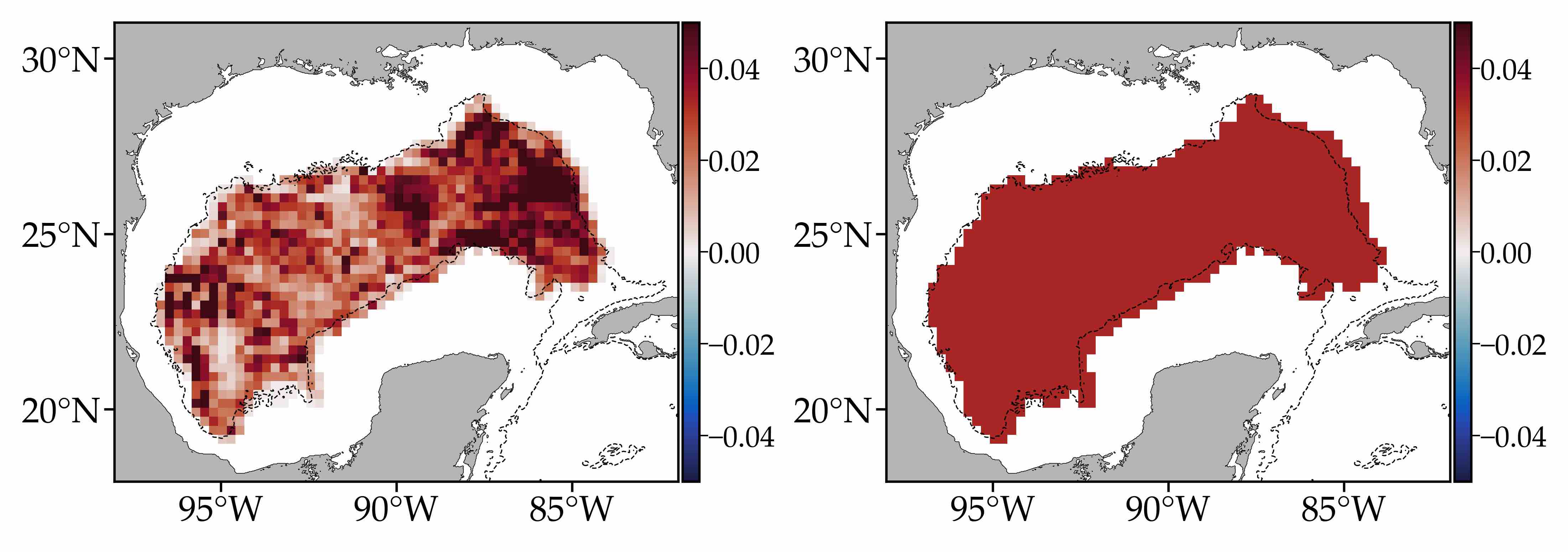}%
  \caption{Left (left panel) and right (right panel) eigenvectors
  associated with the largest eigenvalue of $P$, $\lambda_1 = 1$.
  The left eigenvector is invariant: $pP=p$.  The right eigenvector,
  a vector of ones by row-stochasticity of $P$, is mapped after
  infinitely many applications of $P$ to the left eigenvector
  ($\lim_{k\to\infty} (\mathbf{1}/N)P^k=p$), thereby representing
  a single basin of attraction for the invariant attractor revealed
  by the left eigenvector.}
  \label{fig:eval1}%
\end{figure*}

As expected from the assessment of communication within the Markov
chain associated with $P$, to numerical precision its largest
eigenvalue, $\lambda_1 = 1$, is simple.  The corresponding left and
right eigenvectors are shown in the left and right panels of Fig.\
\ref{fig:eval1}, respectively.  The right eigenvector is flat as
required by row-stochasticity of $P$, which is exactly satisfied
given that no floats exit the domain $X$ on which $P$ is defined
(the physical significance of this domain being closed is discussed
below).  The left eigenvector is positive within the absorbing
closed communicating class of states of the Markov chain, which
covers most of the boxes of the partition (cf.\ Fig.\ \ref{fig:classes}),
and vanishes elsewhere.  Furthermore, the structure of the left
eigenvector resembles quite closely the distribution of the probability
density in the rightmost panel of Fig.\ \ref{fig:limit}.  More
precisely, this is nearly equal to the left eigenvector when
normalized by the area of the boxes in the partition, confirming
its invariant, limiting nature.

\begin{figure*}
  \centering%
  \includegraphics[width=\textwidth]{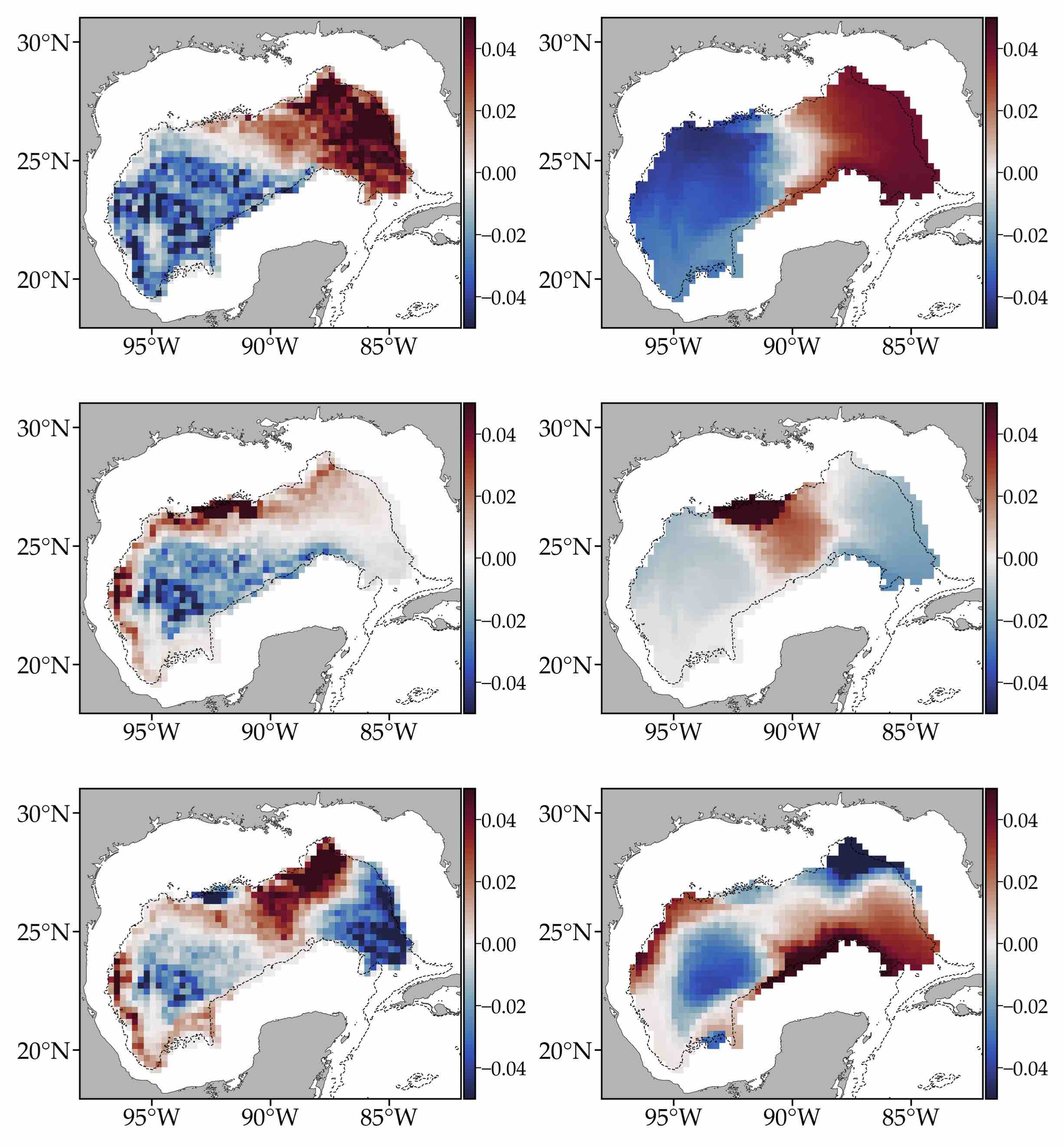}%
  \caption{As in Fig.\ \ref{fig:eval1}, but for $\lambda_2 = 0.9943$
  (top), $\lambda_3 = 0.9822$ (middle), and $\lambda_5 = 0.9701$
  (bottom).  The left eigenvector reveals the locations of
  almost-invariant sets of forward-time attraction. The corresponding
  backward-time basins of attraction are revealed by the right
  eigenvector.  The sign assignment is such that positive (negative)
  right eigenvector portions map to positive (negative) left
  eigenvector portions under repeated left multiplication by $P$.}
  \label{fig:eval235}%
\end{figure*}

The top left and right panels of Fig.\ \ref{fig:eval235} respectively
show the left and right eigenvectors associated with the 2nd largest
eigenvalue of $P$, $\lambda_2 = 0.9943$.  Note that the left
eigenvector takes a single sign within each side of a zero-level
curve that partitions the floats' domain into two regions.  Two
basins of attraction are identified by the right eigenvector.  These
are given by the regions where the right eigenvector is approximately
flat, i.e., approximately looks like $\mathbf 1$.  Splitting the
domain into two regions between which there is weak interaction,
the western (eastern) region constitutes a basin of attraction for
the attractors revealed by the left eigenvector in the western
(eastern) side of the domain.  To make the connection between
forward-time attractors and backward-time basins of attraction
explicit, the eigenvectors have been assigned (in this and the
subsequent figure) signs such that positive (negative) portions of
the right eigenvector map to positive (negative) portions of the
left eigenvector under repeated right multiplication by $P$.  These
attractors are not invariant, but rather will retain tracer for a
finite period of time (intramixing timescales, loosely referred to
as invariance timescales in \citet{Miron-etal-17}, can
be estimated by thinking of $\lambda < 1$ as a decay rate as noted
above; more insightful measures of the invariance time of a set are
provided by the residence and retention times in the set, which are
considered below).

Inspection of the left eigenvector associated with the 3rd largest
eigenvalue of $P$, $\lambda_3 = 0.9822$, reveals further attracting
sets, but with shorter invariance timescale (Fig.\ \ref{fig:eval235},
middle-left panel).  One such set in particular, highlights a
tendency of the Lagrangian motion to circulate along the 1500-m
isobath on the western side of the domain for tracers initially
covering a large sector in the center of the domain.  The latter
is revealed in the right eigenvector (Fig.\ \ref{fig:eval235},
middle-right panel), which is nearly flat in the noted central
sector.  The right eigenvector is also approximately flat on two
regions flanking this sector which are weakly connected through a
very narrow channel running along the southern edge of the domain.
The two regions cover the entire domain, forming two basins of
attraction for almost-invariant sets in the regions of the left
eigenvector with like sign.

Additional almost-invariant attracting sets (with shorter invariance
timescales) and corresponding basins of attraction are revealed by
the left--right eigenvector pairs associated with the 4th to 6th
eigenvalues of $P$ (cf., e.g., the 5th pair in the bottom row of
Fig.\ \ref{fig:eval235}).  Patching together these and the above
basins of attraction the various Lagrangian geographic partitions
shown in Fig.\ \ref{fig:geography} are obtained.

\subsection{Lagrangian geography}

\begin{figure*}
  \centering%
  \includegraphics[width=\textwidth]{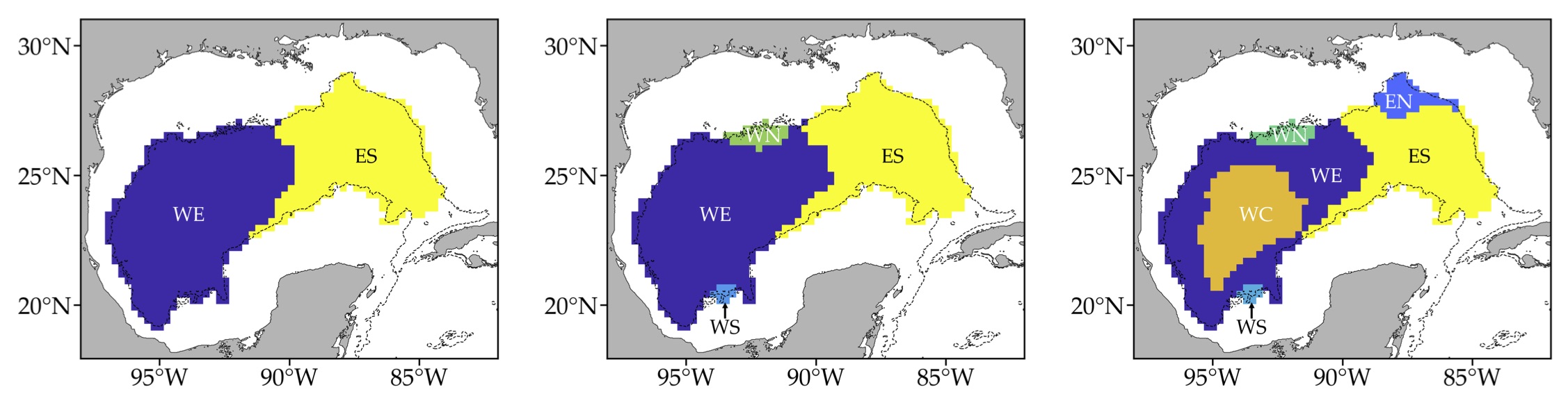}%
  \caption{Lagrangian geography of dynamically weakly interacting
  provinces formed by the domains of attraction associated with the
  most persistent attractors.  Geographic partitions based on the
  analyses of the 2nd right eigenvector, 2nd through 4th right
  eigenvectors, and 2nd through 6th right eigenvectors are shown
  in the left, middle, and right panels, respectively.}
  \label{fig:geography}%
\end{figure*}

Rather than thresholding right eigenvectors as in prior applications
\citep{Froyland-etal-14, Miron-etal-17}, the various provinces in
each Lagrangian geography constructed here were automatically
obtained by applying a $k$-means clustering algorithm
\citep{Kaufman-Rousseeuw-90} that minimizes squared Euclidean
distance as outlined in Algorithm 1 of \citet{Froyland-05}, but
with the weighted fuzzy clustering replaced with $k$-means clustering.
The main geographic partition in the left panel of Fig.\
\ref{fig:geography} was obtained by seeking $k = 2$ clusters in the
2nd right eigenvector of $P$.  Refined geographic partitions, shown
in the middle and right panels, resulted by considering more
eigenvectors in the relevant set suggested by the eigenvalue
uncertainty computation and eigengap inspection. The refined partition
in the right panel of Fig.\ \ref{fig:geography} was obtained by
seeking $k = 6$ clusters in the 2nd through 6th right eigenvector.
The middle panel shows an intermediate partition resulting from
seeking $k = 4$ clusters in the 2nd through 4th right eigenvector.
Seeking $k = 2$ clusters from the 2nd eigenvector is an obvious
choice that follows from direct inspection of this eigenvector.
Seeking $k = K > 2$ clusters from the 2nd through $K$th eigenvectors
assumes that each right eigenvector adds a new province to the
geography at a time.  The silhouette value $-1 \le s \le 1$
\citep{Rousseeuw-87} is a measure of how similar an object is to
its own cluster compared to other clusters, with a large $s$
indicating that the object is well matched to its own cluster and
poorly matched to neighboring clusters.  The mean $s$ equals to
0.91, 0.54 and 0.56 for the clusters identified using $k = 2$, 4,
and 6 when the first 2, 4, and 6 right eigenvectors are taken into
account, respectively.  Larger $k$ in the latter two cases can
produce more consistent clusters, albeit much smaller and hence
with shorter residence or retention times, so we have not considered
them.

In the 2-eigenvector geography 2 large provinces, one western (WE)
and another one eastern (ES), split the domain nearly in half.  The
4-eigenvector geography incorporates two provinces in WE: a small
northern subprovince (WN) and another southern subprovince (WS)
even smaller.  The 6-eigenvector geography incorporates to WE these
same small subprovinces and another, much larger, central subprovince
(WC).  Province ES is not modified by the 4-eigenvector geography,
while the 6-eigenvector geography alters it by the addition of a
small northern subprovince (EN).

As constructed, the provinces of the above Lagrangian geographies
only weakly dynamically interact.  This imposes constraints on
connectivity within the 1500-to-2500-m layer in the GoM.  More
specifically, the communication between any two provinces is
constrained locally by the level of invariance of the attractors
contained within each of them and remotely by that of any attractors
outside of the provinces but sufficiently close to them.

The level of communication among provinces can be assessed by the
computation of forward-time conditional transition probabilities
between pair of provinces.  Over 7 days, the mean inter-province
probability percentages are 98.5, 97.5, and 94.3 for the 2-, 4-,
and 6-eigenvector geographic partitions.  Note that these are high,
indicating weak intra-province dynamical interaction.  Note also
that the percentages decrease as the number of provinces in the
geography increases.  This reflects in part that communication
within large provinces is less constrained than across their
boundaries.  The resulting transition matrices restricted to the
various geographies are not symmetric, revealing the asymmetric
nature of the Lagrangian dynamics in time.  The asymmetry grows
with the number of provinces in the partition, as can be expected.

\begin{figure*}[t]
  \centering%
  \includegraphics[width=\textwidth]{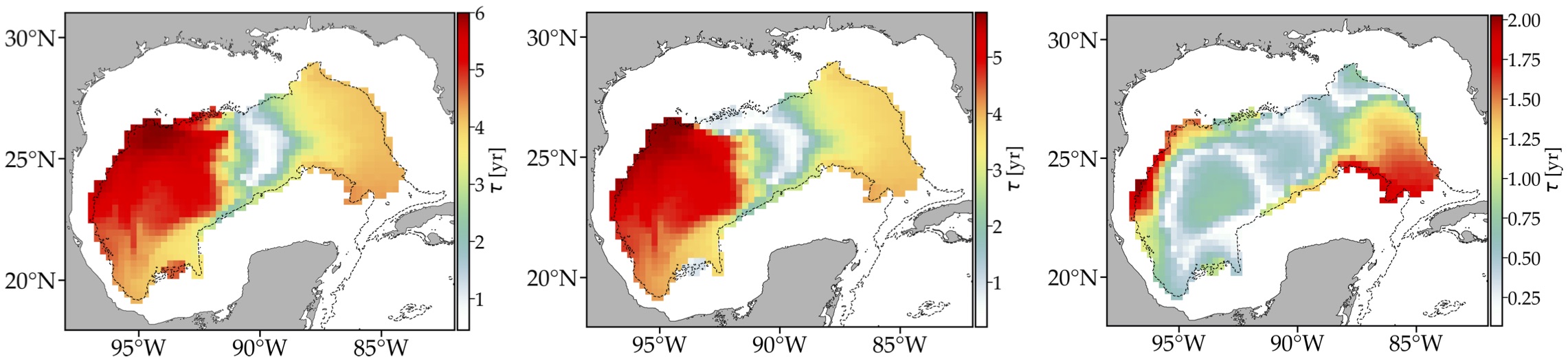}%
  \caption{Estimates of residence time within provinces in the 2-
  (left), 4- (middle), and 6-eigenvectors (right) Lagrangian
  geographies. Note the scale range differences.}
  \label{fig:tau}%
\end{figure*}

From left to right, Figure \ref{fig:tau} shows residence time
estimates according to formula \eqref{eq:res} within each of the
provinces in the 2- to 6-eigenvector Lagrangian geographies.  Note
that $\tau$ decreases outward down to $\tau = 0$ at the boundaries
of the provinces.  As expected, the maximum values tend to be
attained where the right eigenvectors of $P$ locally maximize (in
absolute magnitude).  These regions, as described above, correspond
to basins that are attracted toward forward time almost-invariant
attracting sets.  The small values of $\tau$ attained in the periphery
of the provinces (basins of attraction corresponding to those
attractors) simply indicate that mixing with neighboring provinces
begins at their boundaries.

The residence time calculation shows a west--east asymmetry in the
2-eigenvector geography, with the WE province having longer residence
times than the ES province.  Specifically, the mean residence time
in WE (ES) province is of about 4.53 (3.38) years.  It must be
mentioned that the mean here is taken with respect to a uniform
probability distribution, i.e., it is computed as an average according
to Lebesgue (area) measure.  As noted earlier in the paper, mean
residence times computed using \eqref{eq:res} coincide with retention
times \eqref{eq:ret} based on the likelihood of a trajectory to
survive in a given set if the mean in the former is taken according
to the probability distribution given by the leading left eigenvector
of $P$ restricted to the set in question.  The resulting mean
residence (or, equivalently, retention) times are 4.74 and 3.74
years for the WE and ES provinces, respectively, which are very
similar to the values stated above.

The provinces in the 4- and 6-eigenvector geographies have shorter
residence times.  For instance, on average within the WE, WS, WC,
WW, ES, and EN provinces in the 6-eigenvector partition these are
about 0.74, 0.90, 0.48, 0.64, 1.18, and 0.38 years, respectively.
Shorter residence times in sets covering smaller areas are expected.
But note the short residence time of the WC province despite its
large coverage.

The direct pushforward of tracers with $P$ further revealed that
the slow exchange between the main provinces is executed from east
(west) to west (east) through a northern (southern) corridor.  This
suggests a slow cyclonic circulation in the deep GoM.  Such a
preferred circulatory motion is consistent with the peculiar shape
of the geography in the western side of the domain, which includes
an enclave around which tracers will tend to circulate before
exchange of material is effected.  This explains the residence time
asymmetry of the main provinces.

The west--east residence time asymmetry can be further realized by
computing the time it takes on average to hit or reach a given
province starting in another province.  This can be done using
\eqref{eq:res} with the region $A$ set to the \emph{complement} of
the target province. The result of this calculation for the
6-eigenvector geographic partition is shown in Table \ref{tab:h}.
The top row shows source provinces and the left column target
provinces.  Consider for example the bottom row. The mean time to
hit EN in the eastern basin starting on WC in the western side of
the domain is 6.17 years.  Consider now the second-to-top row.  To
reach WC from EN it takes on average 1.91 years.  Consistent with
west--east residence time asymmetry, it takes more than three times
as long to reach EN from WC.  Clearly, the mean time to reach a
given province starting in the same province is 0.  Note that WS
has not been included in the table as this province is never reached
from outside.

\begin{table}
\caption{Mean time (in years) to reach a province of the 6-eigenvector
  Lagrangian geography indicated in the left column starting from
  any province in the top row.}
\label{tab:h}%
  \centering%
  \begin{tabular}{lccccccc}
	 \hline
       & WE   & WC   & WW   & ES   & EN   \\%
    WE & 0.00 & 0.26 & 0.41 & 0.68 & 0.86 \\%
    WC & 1.05 & 0.00 & 1.46 & 1.73 & 1.91 \\%
    WW & 7.97 & 8.23 & 0.00 & 8.64 & 8.82 \\%
    ES & 1.51 & 1.76 & 1.90 & 0.00 & 0.20 \\%
    EN & 6.17 & 6.42 & 6.56 & 4.68 & 0.00 \\%
	 \hline
  \end{tabular}%
\end{table}

\section{Validation}

\subsection{Chemical tracer}

The deep Lagrangian geography constructed here and the surface
Lagrangian geography computed by \cite{Miron-etal-17} are globally
different on the overlapping domains, suggesting that the surface
Lagrangian motion is to a large extent decoupled from the deep
Lagrangian motion.   An important exception is the partition by a
roughly meridional boundary of the surface and deep domains into
two basins of attraction for almost-invariant attractors revealed
by the inspection of eigenvectors of the corresponding transition
matrices with the 2nd-largest nonunity eigenvalue (compare the left
panel of Fig.\ \ref{fig:geography} and the left panel of Fig.\ 6
of \cite{Miron-etal-17}).

The restricted connection at depth between the eastern and western
GoM was suggested by the behavior of profiling floats parked at
about 900 m launched in the eastern side, which tended to stay
there, and those launched in the western side, which remained there
for a long period of time \citep{Weatherly-etal-05}.  Here we provide
support for the significance of the partition of the deep GoM domain
at a deeper level using the observed evolution of the chemical
tracer injected near the \emph{Deepwater Horizon} oil rig during
the field experiment described by \citet{Ledwell-etal-16}.

\begin{figure*}
  \centering%
  \includegraphics[width=\textwidth]{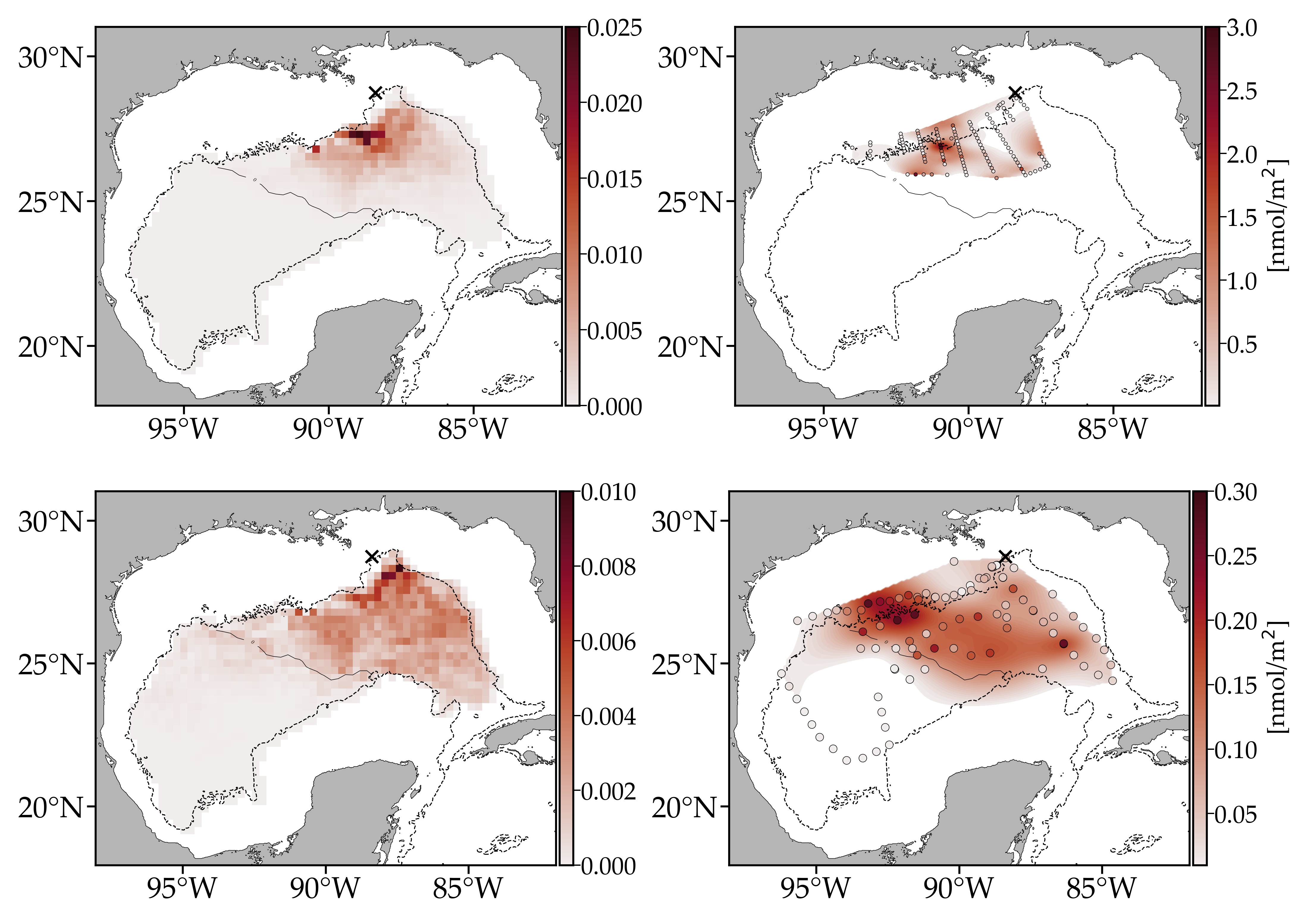}%
  \caption{Comparison between the evolution of a tracer probability
  density under the action of the transition matrix $P$ (left column)
  and dispersion of a chemical tracer (right column) presented both
  after 4 (top row) and 12 (bottom row) months of released about
  100-km southwest of the \emph{Deepwater Horizon} site (indicated
  by a cross). On the right column, the colored background is an
  interpolated map based on column integral of tracer found during
  in-situ casts at the station locations (colored circles). The
  solid line in each panel indicates the zero-level set of the left
  eigenvector of $P$ associated with the 2nd largest eigenvalue.}
  \label{fig:ledwell}%
\end{figure*}

The right panels of Fig.\ \ref{fig:ledwell} show the distribution
taken by the chemical tracer 4 (top) and 12 (bottom) months after
release.  The release site lies about 100-km southwest of the cross,
indicating the location of the \emph{Deepwater Horizon} rig.  The
circles are colored according to the amount of tracer found during
in-situ casts, integrated vertically between 1000 and 2500 m. The
colored background is a smoothed interpolated map based on the
station data.  Note the tendency of the tracer to spread over the
eastern side of the domain.  After 12 months from release, little
tracer is seen to have traversed the zero-level set of the left
eigenvector of $P$ with the largest eigenvalue (indicated by a solid
line).  And note that absolutely no tracer at all was detected
during the in-situ casts made well in the western side of the domain.
This is consistent with the expected fate of a tracer probability
density initiated on the main eastern province of the Lagrangian
geography constructed from the float data.

This expectation is confirmed by the evolution of a tracer probability
started from a source location near the chemical tracer release
site under the action of the transition matrix $P$ (Fig.\
\ref{fig:ledwell}, left panels).  Consistent with the chemical
tracer evolution, the (synthetic) tracer probability spreads similarly
over the eastern side of the domain without significantly crossing
the zero-level set of the largest nontrivial eigenvector.

It must be noted, however, that as the tracer probability is
continually being pushed forward under $P$, it will eventually
spread over the western basin along its northern boundary of the
domain, describing a cyclonic circulatory pattern as noted above
consistent with inferences made from the direct inspection of float
trajectories, the analysis of hydrography and mooring data, and
numerical simulations \citep{Hurlburt-Thompson-80, Oey-Lee-02,
Mizuta-Hogg-04, DeHaan-Sturges-05, Bracco-etal-06, Hamilton-etal-16,
Perez-etal-17, Tenreiro-etal-17}. Also consistent with direct
inspection  of trajectories \citep{Hamilton-etal-16, Perez-etal-17},
and high resolution modelled fields \citep{Bracco-etal-06}, some
probability tracer will circulate cyclonically around the enclave
inside the main western province of the Lagrangian geography.

\subsection{Profiling floats}

\begin{figure*}
  \centering%
  \includegraphics[width=\textwidth]{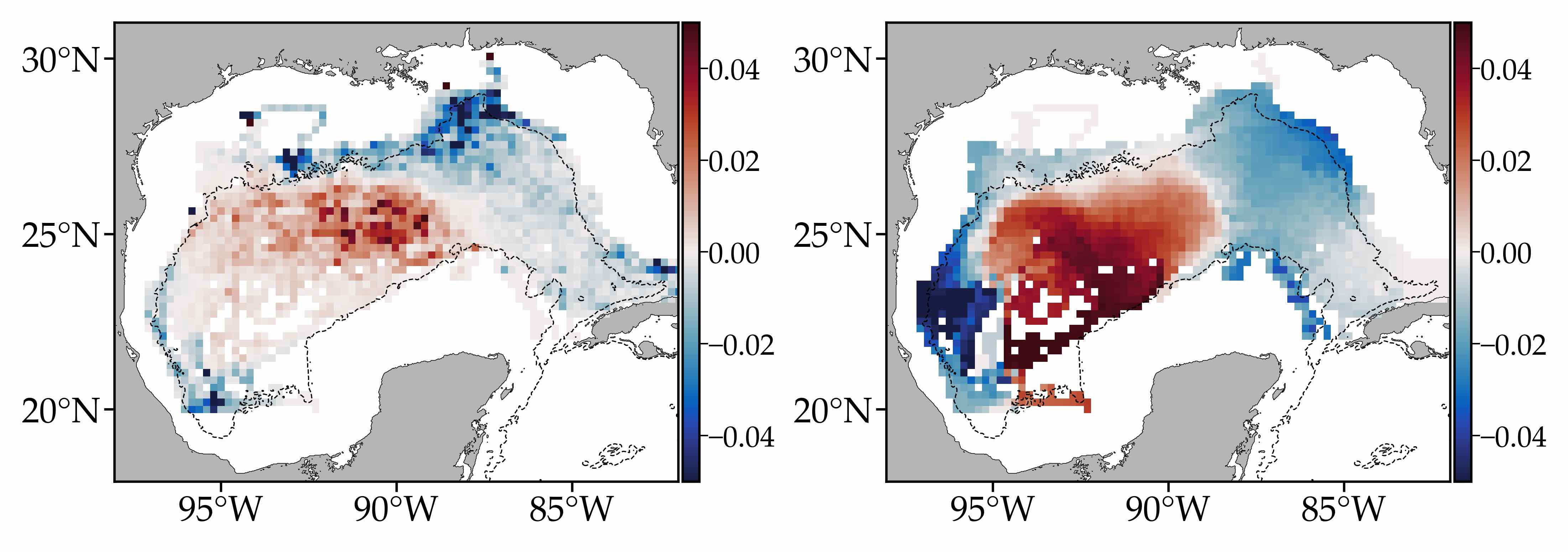}%
  \caption{As in the top row of Fig.\ \ref{fig:eval235}, but based
  on Argo profiling floats.}
  \label{fig:eval1-argo}%
\end{figure*}

Additional independent observational support for the significance
of the results obtained from the analysis of the Markov-chain model
derived using the RAFOS floats is provided by the analysis of a
Markov-chain model constructed using Argo profiling floats drifting
at an average parking depth of $1000$ m.  At a shallower depth, the
Argo trajectories sample a similar horizontal domain of the GoM as
the RAFOS trajectories, but less densely (there are only 60 Argo
floats in the database analyzed).  Also, the temporal coverage of
the Argo floats is not as ample as the RAFOS floats.  With these
differences in mind we constructed a matrix of probabilities of the
Argo floats to transitioning between the boxes of a grid similar
to that used with the RAFOS floats. The transition time was set to
7 days as in RAFOS floats analysis, which required us to interpolate
the original 10-daily trajectories.  The Markov chain associated
with the resulting $P$ was found to be characterized by an absorbing
closed communicating class of states spanning most boxes of the
partition.  Thus $\lambda_1 = 1$ was found to be the only eigenvalue
of $P$ on the unit circle, implying the existence of a limiting
invariant probability vector.  The structure of the left--right
eigenvector pair for the 2nd eigenvalue is shown in Fig.\
\ref{fig:eval1-argo}.  Albeit more noisy and gappy, this pair shows
similarities with that of the $P$ computed using the RAFOS floats
(Fig.\ \ref{fig:eval1}, top row). Indeed, a partition of the domain
nearly into 2 basins of attraction is evident.  This adds confidence
to the RAFOS float analysis.  It also suggests a tendency of the
motion to be preferentially columnar in the deep GoM (recall that
the Argo floats ascend to transmit positions at the surface while
the RAFOS floats remain parked at a fixed level at all times during
an experiment).

\section{Discussion}

\subsection{Cyclonic circulation and \emph{f}/\emph{H}}

\begin{figure*}
  \centering%
  \includegraphics[width=\textwidth]{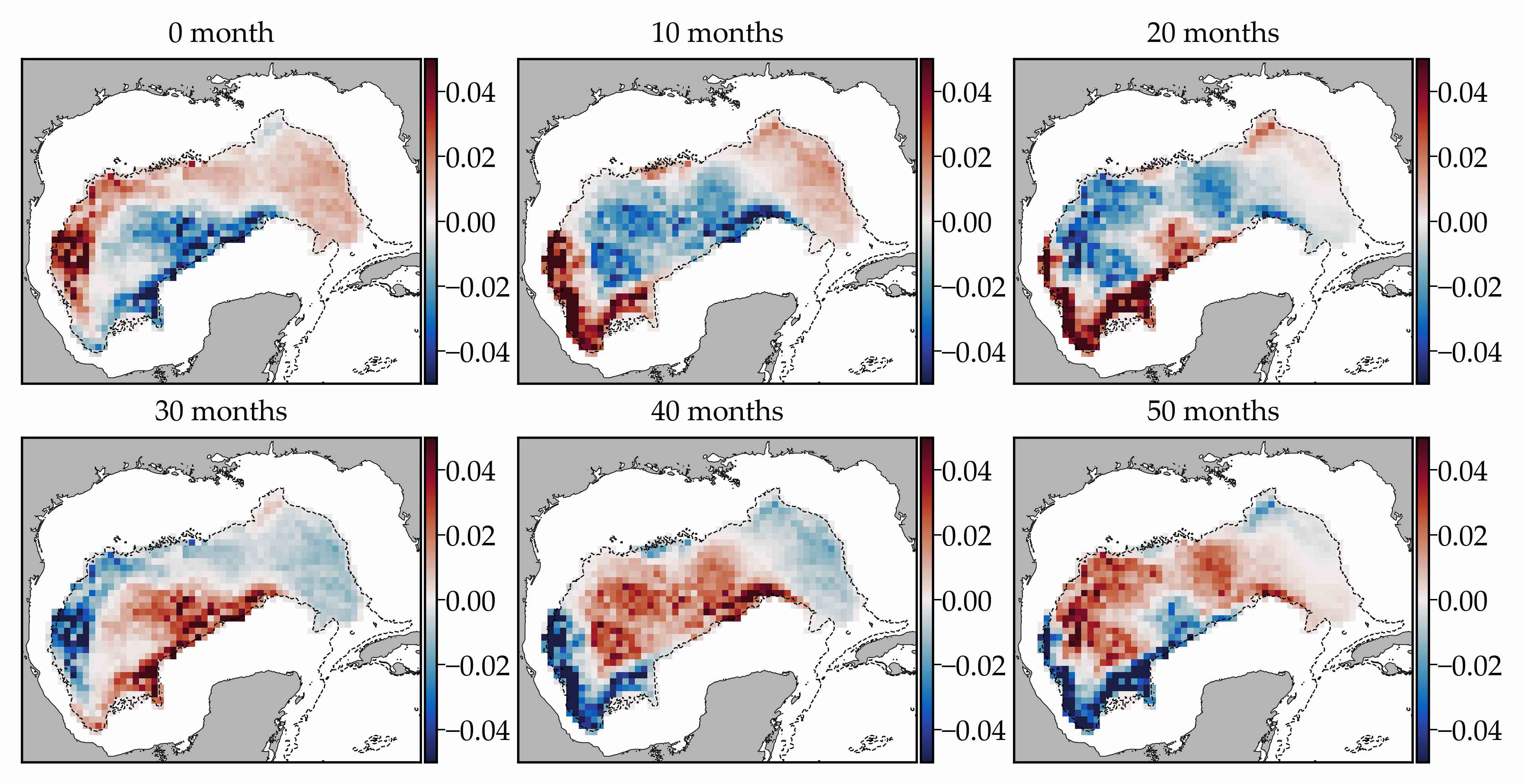}%
  \caption{Snapshots of the forward evolution of the leading complex
  left eigenvector of transition matrix $P$ over an almost cycle
  (with a period of 60.5 months).  The real part is shown.  The
  result does not depend on the member of the complex-conjugate
  eigenvector pair considered.}
  \label{fig:cycle}%
\end{figure*}

The cyclonic circulation in the western side of the GoM domain is
well described by complex eigenvectors of $P$.  Let $v_\pm$ be a
complex-conjugate left eigenvector pair of $P$ with $\lambda_\pm =
r \smash{\e^{\pm\ii \theta}}$, where $\theta \ge 0$.  After $k$
applications of $P$, $v_\pm P^k = r^k v_\pm \smash{\e^{\pm\ii
k\theta}}$.  Viewing each complex component of $v_\pm$ as a vector
in $\mathbb{C}$, if $r \approx 1$ the latter represents a rotation
of each such vectors by an angle $\pm k\theta$.  Furthermore, $v_\pm
P^k$ returns to $v_\pm$ after $k_* = 2\pi/\theta$ applications of
$P$.  Because $k_*$ need not be an integer, $v_\pm P^{k_*}$ will
not in general exactly coincide with $v_\pm$, which represents
almost-cyclic sets.  Figure \ref{fig:cycle} shows $\re{v_\pm P^k}$
where $v_\pm$ corresponds to the leading $\lambda_\pm$ (for which
$r = 0.9775$ and $\theta = 0.0239$) for several $k$ over an almost
cycle (with period $k_*T = 5.0414$ years).  Independent of whether
$v_+$ or $v_-$ is considered, note the cyclonic rotation described
in the western main province of the Lagrangian geography.

Rectification of topographic Rossby waves has been identified as a
driver for the cyclonic circulation along the boundary
\citep{Hurlburt-Thompson-80, Oey-Lee-02, Mizuta-Hogg-04,
DeHaan-Sturges-05}.  In linear, unforced, inviscid, barotropic and
quasigeostrophic dynamics \citep{Gill-82}, the vorticity changes
only when there is motion across the $f/H$ contours, where $f$ is
the Coriolis parameter (twice the local vertical component of the
Earth's angular velocity) and $H$ is the fluid depth. As such, $f/H$
provides a restoring force, supporting topographic Rossby waves,
which through nonlinear interaction can be rectified to give rise
to a mean flow directed mainly along $f/H$ isolines \citep{deVerdiere-79}.
Here we test $f/H$ conservation using the Markov-chain model and
further assess its effect on the evolution of probability tracer
densities in the domain.

Specifically, let $Q_j = f_j/H_i$ be the mean value of $f/H$ inside
box $B_j$ in the partition.  To first-order approximation, $Q_j$
must be preserved along tracer trajectories entering box $B_j$.  To
check this, we propagate backward the observable $f/H$ by $k$ steps
with the transition matrix $P$. Define $q_i := \smash{\sum_{j=1}^N
P_{ij}^kQ_j}$, the $i$th component of the backward propagation of
$Q$, averaged over those boxes $B_j$ that the forward propagation
of $B_i$ intersects.  If $f/H$ were exactly preserved along
trajectories, then one would expect $q_i = Q_i$.

\begin{figure}
  \centering%
  \includegraphics[width=\linewidth]{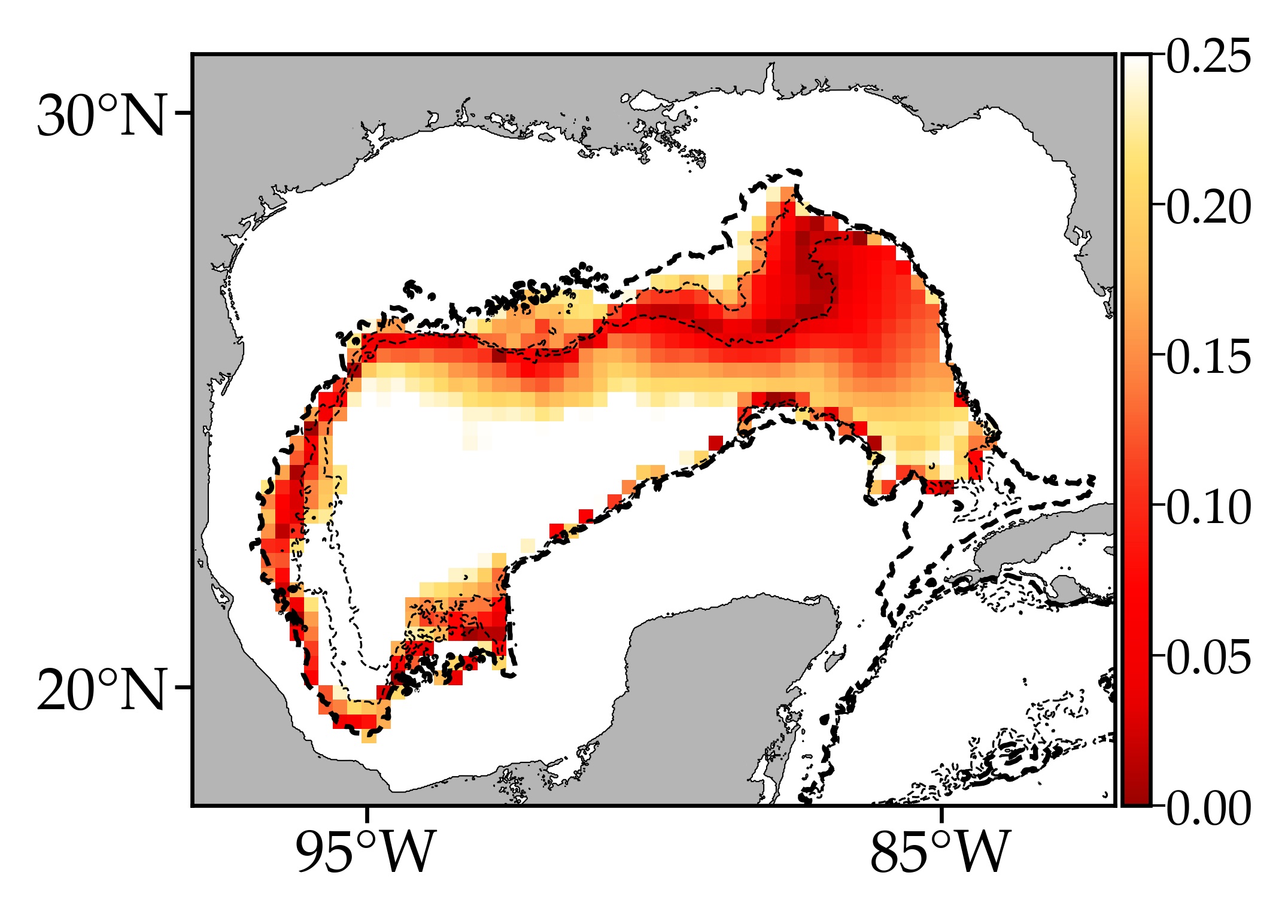}%
  \caption{For each box in the partition, absolute relative error
  between $f/H$, where $f$ is the Coriolis parameter and $H$ is
  depth, and the average of $f/H$ according to the forward evolution
  over 1 year of a probability vector supported in the box.  Dashed
  lines are the isobaths of 1750, 2500, 3000 m.}
  \label{fig:fH}%
\end{figure}

Figure \ref{fig:fH} shows $\varepsilon_i := |q_i/Q_i - 1|$ after
$k = 52$ applications of $P$ (corresponding to 1-year forward
evolution).  Note that $f/H$ is preserved with 25\pct\, error or
less along the western and southern boundaries of the domain and
over a large region of the eastern side of the domain.  Because the
restoring force provided by $f/H$ is largest where $f/H$ varies
rapidly, tracer trajectories initially on the western boundary will
be constrained to run along that boundary as the gradient of $f/H$
across isobaths there is large ($f$ is relatively constant in the
domain).  A similar behavior may be expected for trajectories
starting on the southern boundary across which $f/H$ changes rapidly.
However, $f/H$ is not uniformly preserved along this boundary.
Indeed, boxes where $\varepsilon_i$ are interspersed among boxes
where this is small.  As a consequence, trajectories starting on
that boundary are not expected to be so constrained to run along.
In the eastern portion of the domain where $\varepsilon_i$ is small,
the bottom is relatively flat.  As a result, trajectories starting
will unlikely follow any particular $H$ isoline while nearly
conserving $f/H$.  Moreover, this will tend to occupy the domain
in question, which resembles quite well the positive side of 2nd
left eigenvector of $P$ (cf.\ Fig.\ \ref{fig:eval235}, upper panel).
Finally, in the large western region where $\varepsilon_i$ is large,
the trajectories will wander unrestrained within the region, which
itself resembles quite well the negative side of 2nd left eigenvector
of $P$.

\begin{figure*}
  \centering%
  \includegraphics[width=\textwidth]{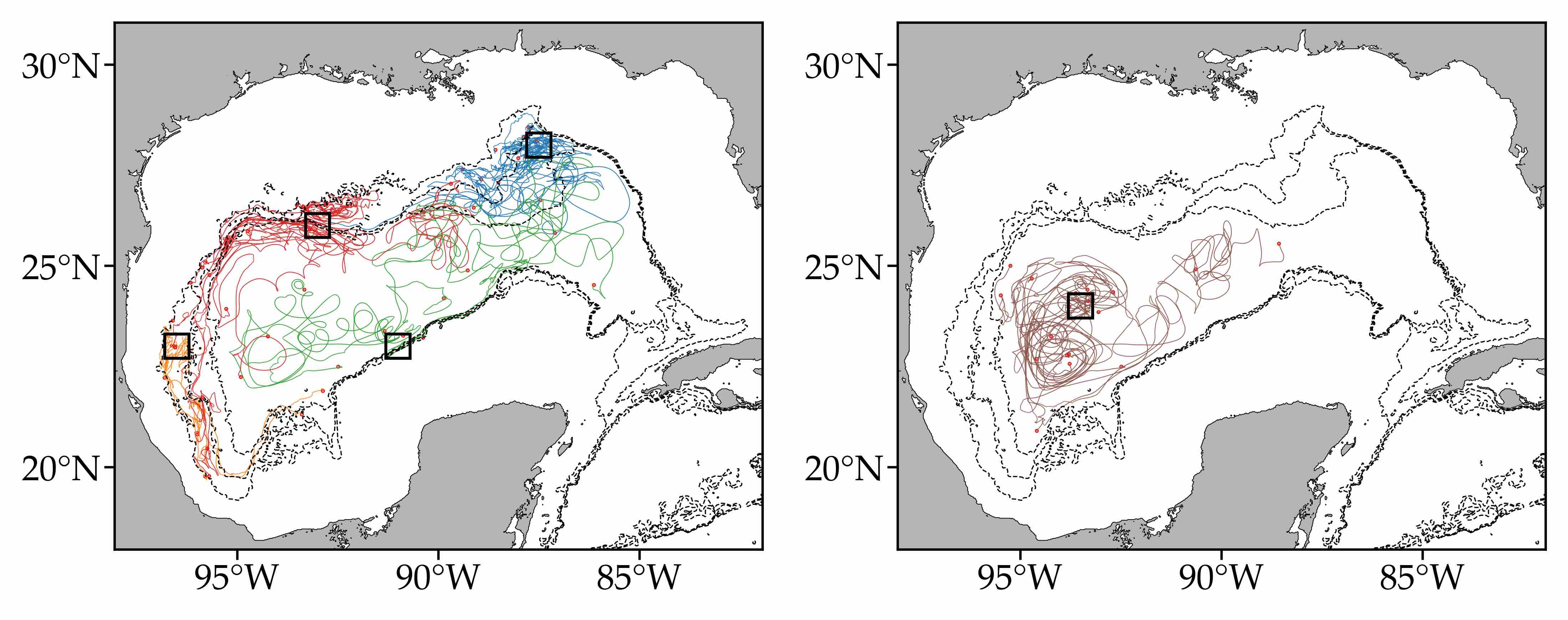}%
  \caption{Trajectories of floats that, having gone through the
  boxes indicated, are constrained by $f/H$ conservation on different
  levels (left) and are not constrained by $f/H$ preservation.}
  \label{fig:fH-floats}%
\end{figure*}

The expected behavior of tracer trajectories deduced from the
Markov-chain model is verified by the behavior observed float
trajectory patterns. This is shown in Fig.\ \ref{fig:fH-floats} for
groups of float trajectories that have gone through selected sites
along the boundary of the domain (left) and the center of the western
side of the domain (right).  Note, in the left panel, how the red
and orange trajectories tend to run along the western boundary,
while the green trajectories are not so constrained to doing so
along the southern boundary.  Observe too in this panel how the
blue trajectories cover the eastern side of the domain consistent
with $f/H$ being preserved in that region.  Finally, note that the
trajectories in the right panel loop around in a largely unrestricted
manner.

\subsection{Homogeneization}

The analysis of the chemical tracer injected at depth suggested
that homogenization in the GoM is more rapid than in the open ocean
\citep{Ledwell-etal-16}.  While the Markov-chain model constructed
here does not predict uniform homogenization in the long run, it
supports a limiting, invariant distribution which does not reveal
a preferred region for accumulation but rather a multitude of
different small regions where some accumulation is possible.  The
highly structured texture of this distribution suggests partial
homogenization in the long run.   This can be quite fast.  For
instance, for a tracer released in the eastern side of the domain,
it can take as short as 1 year or so to spread over that portion
of the domain (cf.\ Fig.\ \ref{fig:tau}) consistent with the good
agreement between the forward evolution of a tracer probability and
the observed chemical tracer spreading (recall Fig.\ \ref{fig:ledwell}).
This corresponds well with the mean time required for the EN province
to hit the WE province, which is of 0.86 years (cf.\ Table \ref{tab:h}
and Fig.\ \ref{fig:geography}).

\subsection{Ventilation}

Below 1000 m the GoM is filled with oxygen-rich water which is
isolated from diffusive inflow of oxygen from the surface by the
presence of a layer of oxygen-poor water \citep{Nowlin-etal-01}.
As standard deep-water formation is very unlikely due to the extreme
cooling and salinity increase required for the surface layer to
sink, ventilation of the deep GoM has been argued to be accomplished
via horizontal transport of oxygen-rich water from the Caribbean
Sea \citep{Rivas-etal-05}.

The tendency of the Lagrangian motion as inferred from the Markov-chain
model constructed using the RAFOS floats to conserve area more
effectively than that using surface drifters, together with the
similarities of the dominant eigenvectors of the transition matrices
built using RAFOS and Argo floats, is consistent with the above
observations in that Lagrangian motion within the 1500--2500-m layer
is predominantly horizontal.  However, the Markov-chain model does
not represent exchanges through the boundary of the domain on which
is supported as the RAFOS floats deployed inside the domain do not
escape the domain.

\begin{figure*}
  \centering%
  \includegraphics[width=\textwidth]{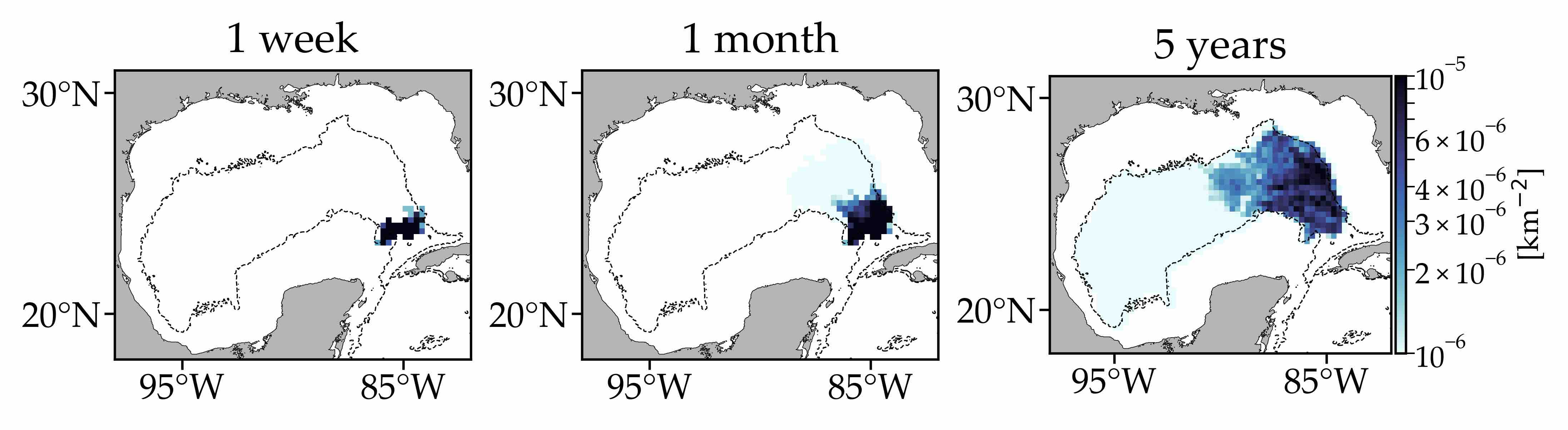}%
  \caption{Snapshots of the evolution under the action of the
  RAFOS-float-based transition matrix of a tracer probability density
  initially in the southeastern corner of the domain.}
  \label{fig:push}%
\end{figure*}

Yet the above does not rule out the possibility that floats deployed
outside the domain eventually enter the domain.  Indirectly, this
possibility is described by our Markov-chain model.  This is shown
in Fig.\ \ref{fig:push} for a tracer probability density initially
at the southeastern corner of the domain.  Note that it spreads
over the domain, mainly cyclonically and along the eastern boundary,
under the action of $P$ as opposed to accumulate in the southeastern
corner.  This suggests a horizontal ventilation pathway from the
Caribbean Sea compensated by weak vertical mixing.

\begin{figure}
  \centering%
  \includegraphics[width=\linewidth]{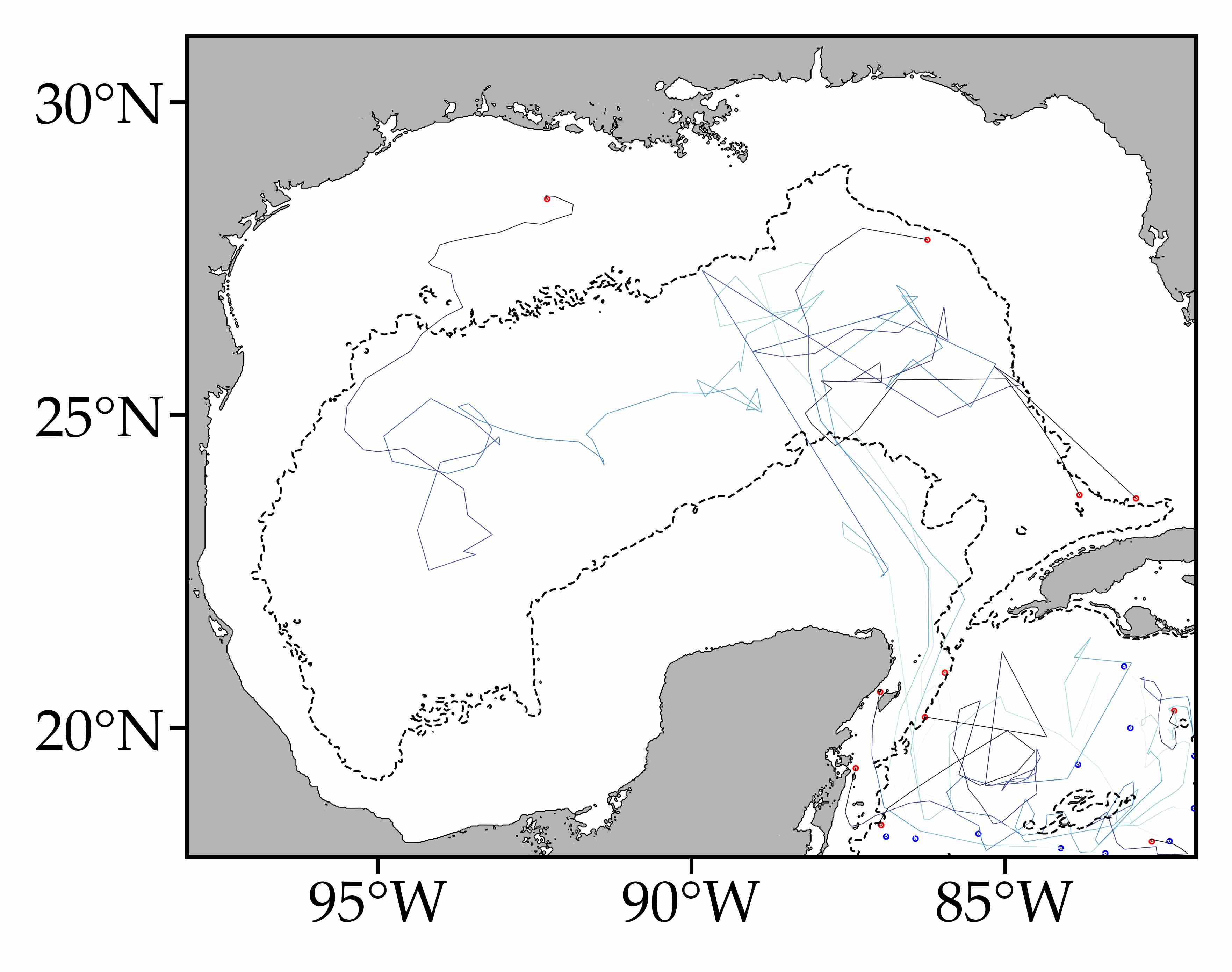}%
  \caption{Trajectories of Argo profiling floats starting inside
  the Caribbean Sea.  Initial positions are indicated by a blue dot
  and final positions by a red dot.}
  \label{fig:argo-inflow}%
\end{figure}

It must be noted that there are no RAFOS floats that can be used
to verify the existence of that pathway.  However, the Argo floats
might suggest it at about 1000 m.  This is shown in Fig.\
\ref{fig:argo-inflow} for a few Argo float trajectories that start
in the Caribbean Sea.  We say ``might'' because we do not know for
certain that the Argo floats penetrate the GoM domain at their
parking depth or at shallower level in their ascend and descend.
Ventilation might well be taken place more effectively at a shallower
level as suggested by the observation that the core of the North
Atlantic Deep Water (NADW), which fills the Caribbean Sea at depth,
located between 1200 and 1300 m \citep{Hamilton-etal-17}.

\subsection{Mean circulation}

\begin{figure*}
  \centering%
  \includegraphics[width=\textwidth]{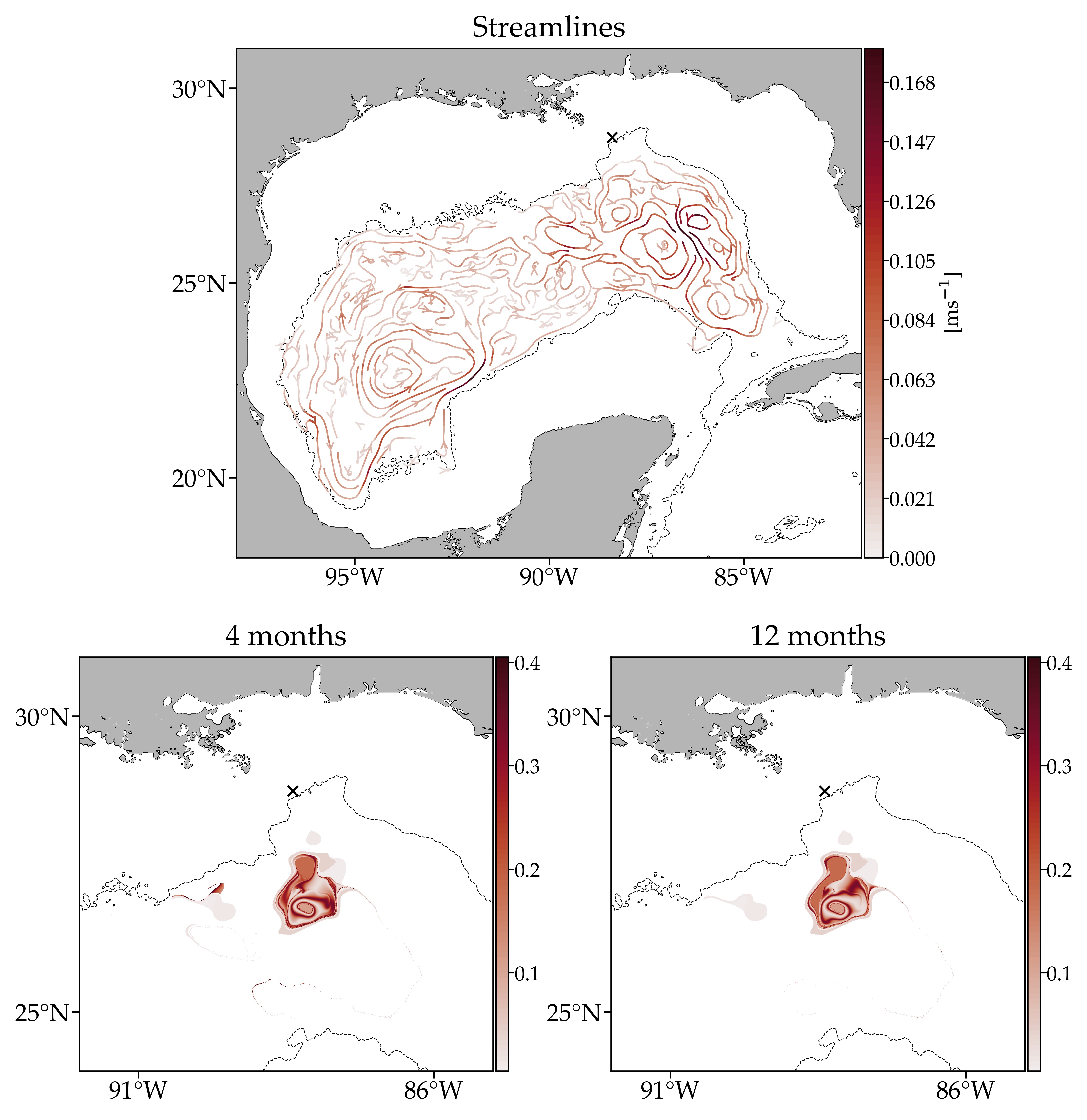}%
  \caption{Ensemble mean streamlines computed using RAFOS float
  velocities (top) and snapshots at 4 (bottom-left) and 12
  (bottom-right) months of the evolution under the corresponding
  flow of a narrow Gaussian initially centered about 100-km southwest
  of the \emph{Deepwater Horizon} oil rig, indicated by a cross.}
  \label{fig:mean}%
\end{figure*}

We close the discussion by showing that the results obtained using
the Markov-chain model could not have been revealed by simply
inspecting the mean circulation deduced from the RAFOS float
trajectories.  The top panel of Fig.\ \ref{fig:mean} shows ensemble-mean
streamlines computed by integrating a steady velocity field resulting
from averaging the float velocities in each box of the grid used
to construct the Markov-chain model.  While the streamlines suggest
a cyclonic flow along the periphery of the domain especially on its
western side, which is consistent with the transfer operator analysis
and also direct inspection of float trajectories \citep{Hamilton-etal-16},
it is difficult to find a correspondence among the many sources and
sinks with the various local minima and maxima of the limiting,
invariant distribution of the Markov-chain model (cf.\ the rightmost
panel of Fig.\ \ref{fig:limit} or the left panel of Fig.\
\ref{fig:eval1}).  The differences with the Markov-chain model
results are most clearly evidenced when the evolution of a tracer
under the corresponding flow is compared with the push forward of
a tracer probability density under the transition matrix $P$.
Snapshots of the evolution of a narrow Gaussian about near the
chemical tracer injection site are shown after 4 and 12 months in
the bottom-left and right panels of Fig.\ \ref{fig:mean}, respectively.
Note that the spread of the tracer is confined to the vicinity of
this site, which is in stark contrast with the much wider spreading
of a probability density source initially near the same location
under action of $P$ or the evolution of the chemical tracer injected
nearby (cf.\ Fig.\ \ref{fig:ledwell}).

\section{Summary and concluding remarks}

Analyzing acoustically tracked (RAFOS) float data in the Gulf of
Mexico (GoM), we have constructed a geography of its Lagrangian
circulation within the deep layer between 1500 and 2500 m revealing
aspects of the circulation transparent to standard Lagrangian data
examination as well as confirming, and thus providing firm support
to, other aspects already noted from direct inspection of the float
trajectories.  The analysis was done by applying a probabilistic
technique that enables the study of long-term behavior in a nonlinear
dynamical system using short-run trajectories.  The Lagrangian
geography is inferred from the inspection of the eigenvectors of a
transfer operator approximated by a transition probability matrix
$P$ of the floats to moving over 1 week between boxes of a grid
laid down on the domain visited by the floats.  Such a transition
matrix provides a Markov-chain representation of the Lagrangian
dynamics.

The basic geography has a single dynamical province which constitutes
the backward-time basin of attraction for a time-asymptotic invariant
attracting set, which is revealed by the unique left eigenvector
of $P$ with unit eigenvalue.  This suggests that the residence time
for tracers within the 1500--2500-m layer is very long.  This result
together with the tendency of the motion to preserve areas more
effectively at depth than at the surface as inferred using
satellite-tracked drifters suggest that transport and mixing is
predominantly lateral.  This is consistent with the idea that
ventilation of the deep Gulf of Mexico is accomplished by the influx
of high-oxygen from the Caribbean Sea through the Yucatan Channel
at depth.  While our Markov-chain model cannot represent exchanges
through its boundary, it indirectly suggests a ventilation conduit
through its southeastern corner balanced by weak vertical mixing.
Some support to this inference was found to be provided by profiling
(Argo) floats deployed in the Caribbean Sea, which were observed
to enter the GoM at depth.

Lateral transport and mixing inside the layer scrutinized does not
happen unrestrainedly.  Indeed, left eigenvectors of $P$ with
eigenvalues close to unity reveal almost invariant sets that attract
Lagrangian tracers originating in disconnected regions where the
corresponding right eigenvectors are nearly flat.  These backward-time
basins of attraction define the provinces of a nontrivial Lagrangian
geography, which, because they are only weakly dynamically interacting,
impose constraints on connectivity and thus on lateral transport
and mixing of tracers.

The simplest nontrivial geographical partition includes 2
nearly equal-area western and eastern provinces.  Tracers initially
released within these main provinces tend to remain confined within
for a few years, with the western province retaining tracers for
longer than the eastern province.  Communication between the provinces
is accomplished through a cyclonic flow confined to the periphery
of the domain, which was shown to be highly constrained by conservation
of $f/H$, where $f$ is the Coriolis parameter and $H$ is depth, in
the western side of the domain.  Smaller secondary provinces of
different shapes with residence times shorter than 1 year or so
were also identified, imposing further restrictions on connectivity
at shorter timescales.

Except for the main provinces, the secondary provinces identified
do not resemble those of the surface Lagrangian geography recently
inferred from satellite-tracked drifter trajectories.  This implies
disparate connectivity characteristics with possible implications
for pollutant (e.g. oil) dispersal at the surface and depth.

The evolution of a chemical tracer from a release experiment as
well as the analysis of a smaller set of Argo floats were shown to
provide independent support for the Lagrangian geography derived
using the RAFOS floats.  It is quite remarkable that the RAFOS and
Argo floats produced similar Markov chain representations of the
Lagrangian dynamics of the deep GoM given the different sampling
characteristics (parking depth, temporal coverage) of these two
observational platforms.

The good agreement between the results from the RAFOS and Argo float
analyses suggests that the probabilistic tools employed here applied
on the global Argo float array may provide important insight into
the abyssal circulation of the world ocean.

\acknowledgments

We thank Alexis Lugo-Fernandez for helping us access the acoustically
tracked float data, which are currently available from the National
Oceanic and Atmospheric Administration (NOAA)/Atlantic Ocean and
Metereological Laboratory (AOML) subsurface float observations page
(http://\allowbreak www.aoml.\allowbreak noaa.gov/\allowbreak
phod/\allowbreak float\_traj/\allowbreak index.php).  The profiling
float data were collected and made freely available through SEA
scieNtific Open data Edition (SEANOE) at http://\allowbreak
www.\allowbreak seanoe.\allowbreak org (doi:10.17882/\allowbreak
42182) by the International Argo Program and the national programs
that contribute to it (http://\allowbreak www.\allowbreak
argo.\allowbreak ucsd.\allowbreak edu,  http://\allowbreak
argo.\allowbreak jcommops.\allowbreak org).  The Argo Program is
part of the Global Ocean Observing System (http://\allowbreak
www.\allowbreak goosocean.\allowbreak org). The chemical tracer
data are publicly available through the Gulf of Mexico Research
Initiative Information \& Data Cooperative (GRIIDC) at https://\allowbreak
data.\allowbreak gulfresearchinitiative.\allowbreak org
(doi:10.7266/\allowbreak N79P2ZK2, doi:10.7266/\allowbreak N75X26VQ,
and doi:10.7266/\allowbreak N7251G4Q).  The surface drifter data
employed in Fig.\ \ref{fig:vertical} involve data publicly available
through the Gulf of Mexico Research Initiative Information and Data
Cooperative (GRIIDC) at https://data.\allowbreak
gulfresearchinitiative.\allowbreak org (doi:10.7266/\allowbreak
N7VD6WC8, doi:10.7266/\allowbreak N7W0940J); NOAA Global Drifter
Program data available at http://www.\allowbreak aoml.noaa.gov/\allowbreak
phod/dac; and data from Horizon Marine Inc.'s
EddyWatch\textsuperscript{\textregistered} program obtained as a
part of a data exchange agreement between Horizon Marine Inc.\ and
Centro de Inevstigaci\'on Cient\'{\i}fica y de Educaci\'on Superior
de Ensenda (CICESE)--Petr\'oleos Mexicanos (Pemex) under PEMEX
contracts SAP-428217896, 428218855, and 428229851.  Support for
this work was provided by the Gulf of Mexico Research Initiative
(PM, FJBV and MJO) as part of CARTHE, the Consejo Nacional de Ciencia
y Tecnolog\'{\i}a (CONACyT)--Secretar\'{\i}a de Energ\'{\i}a (SENER)
grant 201441 (F.J.B.V., M.J.O.\ and P.M.) as part of the Consorcio
de Investigaci\'on del Golfo de M\'exico (CIGoM), and the Australian
Research Council (ARC) Discovery Project DP150100017 (G.F.).

\bibliographystyle{ametsoc2014}

\end{document}